\documentclass[11pt]{article}

\usepackage[T1]{fontenc}
\usepackage[utf8]{inputenc}
\usepackage{lmodern}
\usepackage[a4paper,margin=1in]{geometry}
\usepackage[nopatch=footnote]{microtype}

\usepackage{amsmath,amssymb,amsthm,mathtools,bm}
\numberwithin{equation}{section}

\usepackage{graphicx}
\usepackage{booktabs}
\usepackage{tabularx}
\usepackage{adjustbox}
\usepackage{placeins}
\usepackage{caption}
\usepackage{subcaption}
\usepackage{diagbox}

\usepackage{enumitem}
\usepackage{xcolor}
\usepackage{listings}

\usepackage[hidelinks]{hyperref}
\usepackage[capitalise,noabbrev,nameinlink]{cleveref}

\usepackage{authblk}

\usepackage[backend=biber, style=numeric-comp, sorting=none]{biblatex}
\usepackage{orcidlink}
\newcommand{\orcidRCLSA}{0000-0001-5200-9195} 
\newcommand{\orcidJS}{0000-0002-6016-8011} 

\crefname{equation}{Eq.}{Eqs.}
\Crefname{equation}{Equation}{Equations}
\crefname{figure}{Fig.}{Figs.}
\Crefname{figure}{Figure}{Figures}
\crefname{table}{table}{Tables}
\Crefname{table}{Table}{Tables}
\crefname{section}{Sec.}{Secs.}
\Crefname{section}{Section}{Sections}
\crefname{theorem}{theorem}{theorems}
\Crefname{theorem}{Theorem}{Theorems}

\newcommand{\dd}{\mathop{}\!\text{d}}
\newcommand{\E}{\mathbb{E}}
\newcommand{\Var}{\operatorname{Var}}

\newcommand{\R}{\mathbb{R}}
\newcommand{\Lcal}{\mathcal{L}}
\newcommand{\Pois}{\operatorname{Pois}}
\newcommand{\Normal}{\mathcal{N}}
\newcommand{\pRef}{p_{\text{ref}}}
\newcommand{\qRef}{q_{\boldsymbol{\phi}}}

\usepackage{algorithm}
\usepackage{algpseudocode}

\usepackage{setspace}
\usepackage{booktabs}
\usepackage{amsmath}
\usepackage{threeparttable}
\algnewcommand\algorithmicoptional{\textbf{Optional}}
\algnewcommand\Optional{\item[\algorithmicoptional:]}
\algrenewcommand\algorithmicrequire{\textbf{Require:}}
\algrenewcommand\algorithmicensure{\textbf{Output:}}

\title{Simulation-Based Inference and Unbinned Asimov Construction\\ with Hybrid Neural Density Estimation}

\author[1]{Rafael Coelho Lopes de S\'a \orcidlink{\orcidRCLSA}\,\thanks{Email: \href{mailto:rclsa@umass.edu}{rclsa@umass.edu}}}
\affil[1]{Department of Physics, University of Massachusetts Amherst, MA, USA}

\author[2]{Jay Sandesara \orcidlink{\orcidJS}\,\thanks{Email: \href{mailto:jsandesara@wisc.edu}{jsandesara@wisc.edu}}}
\affil[2]{Data Science Institute, University of Wisconsin--Madison, WI, USA}

\date{\today}

\begin{document}

\maketitle

\begin{abstract}
High-dimensional, unbinned neural simulation-based inference often relies on neural ratio estimation, which uses expressive supervised models to estimate density ratios, but a learned ratio by itself provides neither an explicit normalized density nor a generative model. Flow-based surrogate models instead enable tractable density evaluation and efficient sampling, but residual density-estimation errors can limit precision for complex implicit distributions.

We propose \textit{hybrid neural density estimation}, which uses a flow to define a parameter-independent reference distribution, and classifiers to estimate target-to-reference density ratios.
Multiplying a learned ratio by the reference density gives an evaluable target density surrogate. The ratio also provides importance weights for integration and resampling. We show how this representation defines an exact Asimov dataset, whose maximum likelihood fit returns the generating parameters. We also show how the tractable reference supplies renewable samples for pseudo-experiments and how it enables methods to reduce the Monte Carlo variance of expected test statistic calculations.

We demonstrate the construction in a toy statistical model motivated by high-energy physics measurements, but for which the exact densities are known analytically.

\vspace{0.5cm}
\noindent\textbf{Keywords:} simulation-based inference, neural density estimation, neural ratio estimation, Asimov datasets, neural importance sampling
\end{abstract}

\clearpage
\setcounter{tocdepth}{2}
\tableofcontents

\section{Introduction}
\label{sec:introduction}
Neural simulation-based inference (NSBI) enables likelihood-based analyses of
models specified through simulators, for which observations can be generated
but the likelihood is intractable or expensive to
evaluate~\cite{2020PNAS}. Neural networks learn inference-relevant quantities,
including conditional densities and density
ratios~\cite{Cranmer:2015bka,Brehmer:2018eca,Brehmer:2018hga,Hermans:2020}.
In this work, we propose \textit{hybrid neural density estimation} (hNDE),
which combines a flow-based generative reference with classifier-based
density-ratio estimation. The hNDE approach provides
an evaluable density representation, generation of pseudo-experiments, and
computationally efficient expected sensitivity calculations.

Neural ratio estimation (NRE) and neural density estimation (NDE) offer
complementary capabilities for these tasks. Ratios to a parameter-independent
reference are sufficient for frequentist likelihood ratio-based inference and can be learned
with expressive discriminative models. When the reference resembles the
targets, common features such as sharp peaks, long tails, and detector
resolution effects can largely cancel in the ratio. The remaining function
can then be smoother and have a smaller dynamic range than either absolute
density, making the required precision easier to achieve. The classifier
architectures are not restricted by the normalization, invertibility, or
tractable-Jacobian requirements imposed on many density estimators.

A learned ratio alone, however, supplies neither an absolute density nor a
sampler. Flow-based NDE provides a normalized, evaluable density and a
generative model~\cite{Rezende:2015flows,Papamakarios:2019flows}, but residual
density-estimation errors can affect precision-sensitive inference. In a measurement, two separately trained flows may pass many marginal
distribution checks while their estimated density ratio remains biased by
residual, unrelated errors in the two absolute densities. A hybrid
representation uses the flow to supply the reference density and proposal samples, while classifiers learn target-to-reference ratios for each
target hypothesis.

These capabilities are particularly relevant to unbinned inference in experimental high-energy physics (HEP). Conventional analyses compress reconstructed events into a few binned observables. This construction is robust and
computationally convenient, but compression and binning can lose information
when observables are not sufficient over the full parameter space. The ATLAS
Collaboration developed an NSBI formalism using NRE for parameter
estimation~\cite{ATLAS:2024rpr} and applied it to off-shell Higgs boson
production in the $H^{*}\!\to ZZ\to4\ell$ final
state~\cite{ATLAS:2024tgo}. Its combination of learned ratios with an analytic
factorization of parameter dependence accommodates a vector of parameters of
interest (POIs) and many nuisance parameters (NPs) without densely sampling their complete
joint space during training. The numerical advantages of learning ratios
were central to the precision achieved in these studies.

Subsequent inference steps can nevertheless be expensive. In the ATLAS
implementation, an unbinned Asimov dataset is approximately represented by a large weighted Monte Carlo (MC) sample, and pseudo-experiments are obtained by resampling
that finite pool. Reliable resampling requires the stored pool to be much
larger than the predicted dataset. Processing millions of events in repeated
likelihood evaluations can dominate the cost of profile scans and NP impact
calculations~\cite{ATLAS:2024rpr}. These operational limitations do not
invalidate NRE inference, but make several standard statistical workflows
computationally expensive.

In hNDE, the flow itself defines a parameter-independent
reference, whose support must cover the target distributions. Classifiers are
then trained to distinguish target events from fresh samples drawn from this
flow, thereby estimating target-to-reference density ratios. Multiplying the flow
density by a learned ratio gives an evaluable normalized target-density
surrogate. The ratios also provide importance weights for target-specific expectations and resampling.
In the population limit, an exact ratio recovers the target even when the
flow differs from its original training distribution. This construction
therefore extends ratio based NSBI with density evaluation and renewable
proposal samples, enabling pseudo-experiments from the learned model for
Neyman constructions~\cite{Neyman:1937}.

We apply hNDE to construct weighted unbinned Asimov datasets for expected sensitivity
calculations~\cite{Cowan:2010js}. Fresh events drawn from the learned reference
are weighted to represent the expected event measure at a generating
parameter point. The complete parameter-dependent ratios are normalized on
this same weighted sample, and auxiliary observations are chosen so that their likelihood is maximized at
the generating NPs. These conditions make the generating
point a global maximizer of the finite weighted likelihood. The resulting
construction provides exact internal closure while leaving the accuracy of
the expected likelihood ratio scan to be controlled through numerical
integration and model validation.

The explicit reference density also enables task-specific importance
sampling. Additional independent reference events can be generated on demand,
so numerical integration is no longer limited by a fixed MC pool. We train a
second flow to concentrate quadrature points where they contribute most to
the precision of a chosen Asimov likelihood ratio scan.
Importance weights account for the change of proposal, so direct reference
sampling and the optimized proposal approximate the same population
objective. Combined with the finite Asimov construction, this approach can
reduce the number of events processed in repeated inference calculations.
The reference model, Asimov construction, and importance proposal form the
three algorithms developed in this paper.

We demonstrate the approach in a five-dimensional HEP-inspired model with
known underlying densities, including a systematic uncertainty that affects the shape of the probability distribution. We
compare direct and importance-sampled Asimov scans and assess the learned
model using density-level checks and comparisons of estimator and test
statistic distributions from hNDE-generated and independent simulator-generated
pseudo-experiments at representative parameter points. These comparisons
distinguish numerical consistency within the learned model from agreement
with the simulator. Agreement supports the use of the constructed Asimov
dataset for expected sensitivity calculations and of surrogate-generated
ensembles for inference on the original model within the validated parameter
region.

\Cref{sec:ratioonlynsbi} reviews NRE, and \cref{sec:hybrid} introduces hNDE.
\Cref{sec:lhc-application} develops its application to extended likelihoods,
Asimov datasets, pseudo-experiments, and neural importance sampling (NIS).
The numerical demonstration is presented in \cref{sec:demonstration},
followed by the discussion and conclusions in
\cref{sec:discussion,sec:conclusion}.

\section{Neural ratio estimation for simulation-based inference}
\label{sec:ratioonlynsbi}
In simulation-based inference (SBI), a simulator generates observations
$\mathbf{x}\sim p(\cdot;\boldsymbol{\theta})$, while the density itself
is intractable or expensive to evaluate. Neural models trained on
simulated observations can approximate densities or density ratios
directly in the observable space. 

We will denote
$\boldsymbol{\theta}=(\boldsymbol{\mu},\boldsymbol{\alpha})$, where
$\boldsymbol{\mu}$ denotes the POIs and
$\boldsymbol{\alpha}$ the NPs. For the observed data, we denote the likelihood by
$\Lcal(\boldsymbol{\theta})$. The frequentist objective considered
here is the profile likelihood ratio statistic
\begin{equation}
  t_{\boldsymbol{\mu}}
  =-2\log
  \frac{\Lcal(\boldsymbol{\mu},
    \widehat{\widehat{\boldsymbol{\alpha}}}_{\boldsymbol{\mu}})}
       {\Lcal(\widehat{\boldsymbol{\mu}},
    \widehat{\boldsymbol{\alpha}})},
  \label{eq:full_density_model}
\end{equation}
which is used for hypothesis tests and confidence regions.
The unconditional and conditional maximum-likelihood estimators are
\begin{equation}
  (\widehat{\boldsymbol{\mu}},\widehat{\boldsymbol{\alpha}})
  =\underset{\boldsymbol{\mu},\boldsymbol{\alpha}}
    {\operatorname{argmax}}\,
    \Lcal(\boldsymbol{\mu},\boldsymbol{\alpha}),
  \qquad
  \widehat{\widehat{\boldsymbol{\alpha}}}_{\boldsymbol{\mu}}
  =\underset{\boldsymbol{\alpha}}{\operatorname{argmax}}\,
    \Lcal(\boldsymbol{\mu},\boldsymbol{\alpha}).
  \label{eq:general-mle-definitions}
\end{equation}

Two common surrogate targets are the density and a density ratio,
\begin{equation}
  q_\psi(\mathbf{x};\boldsymbol{\theta})
  \simeq p(\mathbf{x};\boldsymbol{\theta}),
  \qquad
  r_{\boldsymbol{\psi}}(\mathbf{x};\boldsymbol{\theta})
  \simeq
  \frac{p(\mathbf{x};\boldsymbol{\theta})}{\pRef(\mathbf{x})},
\end{equation}
where $\pRef$ is a parameter-independent reference density whose
support covers the target family. Learning the conditional density
uses NDE, also called neural likelihood
estimation in this setting. Learning the ratio uses NRE.

Both NDE and NRE can estimate a density or density ratio for the
complete dataset. For a fixed number $n$ of independent and identically distributed (iid) events, either
method can instead learn the corresponding quantity for an
individual event. The dataset density and its ratio to the product
reference density are then approximated by
\begin{equation}
  \prod_{i=1}^{n}q_\psi(\mathbf{x}_i;\boldsymbol{\theta}),
  \qquad
  \prod_{i=1}^{n}r_{\boldsymbol{\psi}}
    (\mathbf{x}_i;\boldsymbol{\theta}),
\end{equation}
respectively. This reduces the dimensionality of the learning
problem for both approaches. The HEP application in
\cref{sec:lhc-profile} exploits this structure, with event
counts and auxiliary measurements included in the extended
likelihood.

\subsection{Classifiers as density-ratio estimators}
\label{sec:ratio}

A classifier $c_{\boldsymbol{\psi}}(\mathbf{x};\boldsymbol{\theta})$
is trained to distinguish target observations drawn from
$p(\cdot;\boldsymbol{\theta})$ from reference observations drawn
from $\pRef$. For a parameterized classifier, the parameters are
sampled from the same distribution in both classes.
With balanced training classes, the population optimum of the
binary cross-entropy loss is
\begin{equation}
  c^\star(\mathbf{x};\boldsymbol{\theta})
  =\frac{p(\mathbf{x};\boldsymbol{\theta})}
         {p(\mathbf{x};\boldsymbol{\theta})+\pRef(\mathbf{x})}.
\end{equation}
The trained classifier therefore provides the ratio
estimate~\cite{Cranmer:2015bka}
\begin{equation}
  r_{\boldsymbol{\psi}}(\mathbf{x};\boldsymbol{\theta})
  \propto\frac{c_{\boldsymbol{\psi}}(\mathbf{x};\boldsymbol{\theta})}
         {1-c_{\boldsymbol{\psi}}(\mathbf{x};\boldsymbol{\theta})}
  \simeq
  \frac{p(\mathbf{x};\boldsymbol{\theta})}{\pRef(\mathbf{x})}.
  \label{eq:classifier-ratio}
\end{equation}
The overall scale is absorbed into $r_{\boldsymbol{\psi}}$ so that its
mean under the reference distribution is one, as made explicit for hNDE
in \cref{eq:hybrid-population-normalization}.
The ratio is exact when $c_{\boldsymbol{\psi}}=c^\star$.
For unbalanced classes, the odds must also be corrected for the
class proportions used in training. Calibration, reweighting
closure, ensemble stability, and checks of fitted parameters
against independent simulation are used to assess the accuracy
of the learned ratios~\cite{ATLAS:2024rpr}.%

\section{Hybrid neural density estimation}
\label{sec:hybrid}
A normalized flow density trained on a representative distribution
$p_{\mathrm{ref}}(\mathbf{x})$ is denoted by $\qRef(\mathbf{x})$.
Once trained, the flow itself defines the reference hypothesis. A balanced
parameterized classifier is then trained to distinguish target events from
fresh samples drawn from $\qRef$, using the same parameter distribution for
both classes. For a target family $p(\mathbf{x};\boldsymbol{\theta})$, the
classifier estimates a target-to-reference density ratio and defines a hybrid
density surrogate,
\begin{equation}
  r_{\boldsymbol{\psi}}(\mathbf{x};\boldsymbol{\theta})
  \simeq \frac{p(\mathbf{x};\boldsymbol{\theta})}{\qRef(\mathbf{x})},
  \qquad
  \widehat p_{\mathrm{hNDE}}(\mathbf{x};\boldsymbol{\theta})
  =\qRef(\mathbf{x})r_{\boldsymbol{\psi}}(\mathbf{x};\boldsymbol{\theta}).
  \label{eq:hybrid-density}
\end{equation}

For an exact ratio, this product recovers the target even when
$\qRef\neq p_{\mathrm{ref}}$. The flow need not independently provide a
precise model of each target. The reference must be evaluable, inexpensive to
sample, and nonzero wherever a target density is nonzero. Regions where the target is appreciable but the reference is very small can produce large ratios, making classifier training and subsequent importance sampling less efficient.

Throughout, $r_{\boldsymbol{\psi}}$ denotes the population-normalized
learned ratio, with its overall scale absorbed in the definition:
\begin{equation}
  \E_{\qRef}[r_{\boldsymbol{\psi}}(\cdot;\boldsymbol{\theta})]=1
  \qquad\text{for every }\boldsymbol{\theta}.
  \label{eq:hybrid-population-normalization}
\end{equation}
Thus \cref{eq:hybrid-density} defines a normalized continuous surrogate.
Any overall scale of the classifier odds cancels from the finite-sample
normalized ratio below, so it need not be evaluated separately.

In practice, expectations under $\qRef$ are approximated using a fixed
weighted reference sample
$X_M=\{\mathbf{x}_m,\omega_m\}_{m=1}^M$, with $\omega_m\geq0$ and
$\sum_m\omega_m=1$. For independent draws from $\qRef$, $\omega_m=1/M$.
We define

\begin{equation}
  Z(\boldsymbol{\theta})
  =\sum_{m=1}^{M}\omega_m
  r_{\boldsymbol{\psi}}(\mathbf{x}_m;\boldsymbol{\theta}),
  \qquad
  \widetilde r_{\boldsymbol{\psi}}(\mathbf{x};\boldsymbol{\theta})
  =\frac{r_{\boldsymbol{\psi}}(\mathbf{x};\boldsymbol{\theta})}
  {Z(\boldsymbol{\theta})},
  \label{eq:empirical-ratio-normalization}
\end{equation}
where $Z>0$ is required. This ratio depends on the chosen sample
and satisfies
$\sum_m\omega_m\widetilde r_{\boldsymbol{\psi}}
(\mathbf{x}_m;\boldsymbol{\theta})=1$ exactly. The identity normalizes the
finite sample, while $Z$ converges to one as the reference quadrature
converges. This procedure removes a finite-sample normalization offset, but does not
repair local classifier bias. Calibration, ensemble stability,
multidimensional reweighting, and inference-level closure against independent
simulator samples remain necessary. On this sample, the learned target is represented by the probabilities
\begin{equation}
  w_m(\boldsymbol{\theta})
  =\omega_m\widetilde r_{\boldsymbol{\psi}}
  (\mathbf{x}_m;\boldsymbol{\theta}),
  \qquad
  \sum_m w_m(\boldsymbol{\theta})=1.
  \label{eq:target-weights}
\end{equation}
Weighted averages approximate expectations under the continuous surrogate,
while resampling indices with probabilities $w_m$ generates events from its
finite empirical approximation. Additional independent reference points can
be generated from the flow to improve this approximation. The training
procedure is summarized in Algorithm~\ref{alg:hNDE}.

\begin{algorithm}[tbp!]
\setstretch{0.9}
\caption{hybrid Neural Density Estimation (hNDE)}
\label{alg:hNDE}
\begin{algorithmic}

\Require Simulator $\mathbf{x}\sim p(\cdot;\boldsymbol{\theta})$;
training proposal $\rho_{\mathrm{train}}(\boldsymbol{\theta})$;
reference sampler $\mathbf{x}\sim p_{\mathrm{ref}}$;
reference-flow sample size $N_{\mathrm{flow}}$;
ratio-training sample size $N_{\mathrm{ratio}}$

\Ensure Frozen model $\{\qRef,r_{\boldsymbol{\psi}}\}$ and evaluable
normalized surrogate
$\widehat p_{\mathrm{hNDE}}(\mathbf{x};\boldsymbol{\theta})$
\vspace{0.5em}

\Statex \textit{Phase 1: reference flow}
\State Draw $\mathbf{x}^{\mathrm{ref}}_i\sim p_{\mathrm{ref}}$,
$i=1,\ldots,N_{\mathrm{flow}}$
\State Train $\qRef$ by maximizing
\[
  \sum_{i=1}^{N_{\mathrm{flow}}}
  \log\qRef(\mathbf{x}^{\mathrm{ref}}_i)
\]
\State Freeze the flow parameters; $\qRef$ defines the reference used below

\Statex \textit{Phase 2: target-to-reference ratios}
\State Draw matched training pairs, for $i=1,\ldots,N_{\mathrm{ratio}}$,
\[
  \boldsymbol{\theta}_i\sim\rho_{\mathrm{train}},\qquad
  \mathbf{x}_i^+\sim p(\cdot;\boldsymbol{\theta}_i),\qquad
  \mathbf{x}_i^-\sim\qRef
\]
\State Train a balanced classifier
$c_{\boldsymbol{\psi}}(\mathbf{x};\boldsymbol{\theta})$ to distinguish
$(\mathbf{x}_i^+,\boldsymbol{\theta}_i)$ from
$(\mathbf{x}_i^-,\boldsymbol{\theta}_i)$
\State Freeze the classifier parameters and define the population-normalized ratio
\[
  r_{\boldsymbol{\psi}}(\mathbf{x};\boldsymbol{\theta})
  \propto\frac{c_{\boldsymbol{\psi}}(\mathbf{x};\boldsymbol{\theta})}
  {1-c_{\boldsymbol{\psi}}(\mathbf{x};\boldsymbol{\theta})},
  \qquad \E_{\qRef}[r_{\boldsymbol{\psi}}(\cdot;\boldsymbol{\theta})]=1
\]

\Statex \textit{Phase 3: hybrid density surrogate}
\State Form
\[
  \widehat p_{\mathrm{hNDE}}(\mathbf{x};\boldsymbol{\theta})
  =\qRef(\mathbf{x})r_{\boldsymbol{\psi}}(\mathbf{x};\boldsymbol{\theta})
\]
\State \Return $\{\qRef,r_{\boldsymbol{\psi}}\}$ and
$\widehat p_{\mathrm{hNDE}}$; finite sample normalization is
defined in \cref{eq:empirical-ratio-normalization,eq:target-weights}

\end{algorithmic}
\end{algorithm}

\section{Application to frequentist analyses}
\label{sec:lhc-application}
In this section, we discuss the application of hNDE to frequentist statistical data analyses. In~\cref{sec:lhc-profile}, we consider the case of datasets with a Poisson-distributed event count and
conditionally iid event observables. This
\textit{extended likelihood} setting is common in experimental HEP analyses and allows the neural models to be trained at the event level. This formalism will be used for the numerical demonstration in~\cref{sec:demonstration}.

After considering this specific case, we discuss general applications of the hNDE approach to the construction of Asimov datasets. We extend the results from Ref.~\cite{LopesDeSa:finite-asimov} by studying the statistical properties of the unbinned Asimov likelihood and show how the hNDE approach allows for more efficient constructions of the Asimov dataset.

\subsection{Extended likelihoods}
\label{sec:lhc-profile}
For a detector reconstructed iid event observable $\mathbf{x}\in\R^d$ and parameters
$\boldsymbol{\theta}=(\boldsymbol{\mu},\boldsymbol{\alpha})$, comprising
POIs and NPs, the event intensity is
\begin{equation}
  \nu(\mathbf{x};\boldsymbol{\theta})
  =\lambda(\boldsymbol{\theta})p(\mathbf{x};\boldsymbol{\theta}),
  \qquad
  \lambda(\boldsymbol{\theta})
  =\int\nu(\mathbf{x};\boldsymbol{\theta})\dd\mathbf{x}.
  \label{eq:intensity}
\end{equation}
Here $p$ is normalized to unity and $\lambda$ is the expected event yield.
The observed count follows $n\sim\Pois(\lambda(\boldsymbol{\theta}))$.
Conditional on $n$, the events are independent draws from
$p(\cdot;\boldsymbol{\theta})$.

When the model is a nonnegative mixture of physics processes indexed by $s$,
\begin{equation}
  \nu(\mathbf{x};\boldsymbol{\theta})
  =\sum_s\lambda_s(\boldsymbol{\theta})p_s(\mathbf{x};\boldsymbol{\theta}),
  \qquad
  \lambda(\boldsymbol{\theta})=\sum_s\lambda_s(\boldsymbol{\theta}),
  \label{eq:sample-intensity}
\end{equation}
where each $p_s$ is normalized and $\lambda_s\geq0$ is its expected yield.
The extended likelihood also applies to a total intensity specified without
this decomposition.

In HEP applications, the index $s$ commonly labels distinct simulated physics
samples that contribute to the observed event population. Their expected
yields and density ratios can therefore be supplied and combined process by
process.

Up to parameter-independent factors, the likelihood can be written as
\begin{equation}
  \Lcal(\boldsymbol{\theta})
  =e^{-\lambda(\boldsymbol{\theta})}
  \prod_{i=1}^{n}\nu(\mathbf{x}_i;\boldsymbol{\theta})\,
  f(\mathbf{a};\boldsymbol{\alpha}),
  \label{eq:extended-likelihood}
\end{equation}
where $f$ is the likelihood of auxiliary observations $\mathbf{a}$,
assumed independent of the event sample conditional on the parameters.
For independent constraints on individual NPs,
$f(\mathbf{a};\boldsymbol{\alpha})=\prod_k f_k(a_k;\alpha_k)$;
correlated auxiliary measurements can instead be described by a joint $f$.
The profile likelihood-ratio statistic is defined in \cref{eq:full_density_model}
with the estimators defined in \cref{eq:general-mle-definitions}, using
the full extended likelihood.

The auxiliary observables $\mathbf{a}$ are often called global observables
in HEP likelihood implementations. Their constraint likelihood encodes the
information supplied by auxiliary measurements that constrain the NPs.

In frequentist applications of NRE, the event density can be represented by a ratio to an
implicit, parameter-independent reference $\pRef$, either for the complete
process mixture or separately for each simulated sample. The contribution
$\sum_i\log\pRef(\mathbf{x}_i)$ cancels from profile log-likelihood ratios,
so the reference density need not be evaluated for inference. This ratio
representation is sufficient for likelihood-based inference, but does not
by itself provide an evaluable normalized density or a direct sampler.
Two operational consequences are particularly relevant to HEP analyses.

First, expected sensitivity is conventionally evaluated using an Asimov
dataset, a representative dataset whose fit returns the generating parameters
and which encodes the median expected sensitivity under the usual asymptotic
assumptions~\cite{Cowan:2010js}. In an unbinned NRE analysis, a large weighted
MC sample approximates this dataset. Obtaining sufficient integration
precision can require repeated evaluations of the neural networks on millions
of events during profile scans and NP impact decompositions. Likelihood
maximizations requiring hundreds of gigabytes of memory and hours of wall-clock time
have been reported for the ATLAS implementation~\cite{ATLAS:2024rpr}.

Second, a Neyman construction requires pseudo-experiment ensembles at the
tested parameter points, while the trained ratio model alone supplies no
direct sampler. A larger reference pool can be resampled with probabilities
proportional to event weights or learned ratios, but the stored pool remains
the resolution limit of the generator. In the ATLAS procedure, each reference
event is assigned a Poisson-distributed integer multiplicity whose mean equals
its Asimov weight~\cite{ATLAS:2024rpr}. The hNDE construction addresses these
limitations by replacing the implicit reference with a normalized density
that can be evaluated and sampled, while retaining the same ratio
representation.

For hNDE, each process is represented by a process-to-reference ratio,
with the same reference flow used for all processes:
\begin{equation}
  r_{s,\boldsymbol{\psi}}(\mathbf{x};\boldsymbol{\theta})
  \simeq\frac{p_s(\mathbf{x};\boldsymbol{\theta})}{\qRef(\mathbf{x})},
  \qquad
  \E_{\qRef}[r_{s,\boldsymbol{\psi}}(\cdot;\boldsymbol{\theta})]=1.
  \label{eq:hybrid-sample-density}
\end{equation}
Replacing each $p_s$ by its normalized surrogate
$\qRef r_{s,\boldsymbol{\psi}}$ defines the learned intensity, for
which we retain the notation $\nu$:
\begin{equation}
  h(\mathbf{x};\boldsymbol{\theta})
  =\sum_s\lambda_s(\boldsymbol{\theta})
    r_{s,\boldsymbol{\psi}}(\mathbf{x};\boldsymbol{\theta}),
  \qquad
  \nu(\mathbf{x};\boldsymbol{\theta})
  =\qRef(\mathbf{x})h(\mathbf{x};\boldsymbol{\theta}).
  \label{eq:h-definition}
\end{equation}
The factor $\qRef$ cancels from profile likelihood ratios. The function $h(\mathbf{x},\boldsymbol{\theta})$ is used to convert between the reference density and the intensity $\nu(\mathbf{x},\boldsymbol{\theta})$. The yields and parameter-dependent ratios remain part of the likelihood.

Consequently, the common flow factor need not be evaluated in each profile
likelihood calculation, while it remains available for absolute density
evaluation and sample generation. A single reference must cover every physics
sample and parameter point used for inference. In a HEP analysis, efficient
ratio estimation is aided by a reference that reproduces the resonant
structure, tails, and detector effects present in the physical samples.

The ratios can be parameterized directly or assembled from nominal
process-to-reference ratios and ratios for systematic shape variations,
together with analytic yield and shape interpolation functions. The latter
approach follows the ATLAS NSBI formalism~\cite{ATLAS:2024rpr} and matches
the way systematic variations are commonly supplied in HEP analyses.

Separate training of the nominal process-to-reference ratios and systematic shape ratios\footnote{For each process $s$, \textit{yield} denotes the expected
number of events $\lambda_s(\boldsymbol{\theta})$, while \textit{shape}
denotes the normalized distribution of the observables,
$p_s(\mathbf{x};\boldsymbol{\theta})$.}
allows the analytic model to combine these ingredients with yield and shape
interpolation functions. This decomposition scales to statistical models
with many NPs. When an analytic factorization is unavailable,
the complete dependence of $p_s(\mathbf{x};\boldsymbol{\theta})$ can instead
be learned by a parameterized surrogate~\cite{Brehmer:2018eca,Brehmer:2018hga}.

At a generating point $\boldsymbol{\theta}$, we prepare a large
reference pool of events $\{\mathbf{x}_m,\omega_m\}_{m=1}^{M}$, where $\omega_m=1/M$. The probability of selecting
reference event $m$ for process $s$ is
$w_m^{(s)}(\boldsymbol{\theta})=\omega_m r_{s,\boldsymbol{\psi}}
(\mathbf{x}_m;\boldsymbol{\theta})=r_{s,\boldsymbol{\psi}}/M$. For nonnegative component yields, draw independent counts
$n_s\sim\Pois(\lambda_s(\boldsymbol{\theta}))$ and, for each process,
resample $n_s$ reference indices independently with replacement using
the probabilities $w_m^{(s)}(\boldsymbol{\theta})$. Each selected occurrence
enters the pseudo-experiment with unit weight, including repetitions. In practice, the number of events sampled from $\qRef$ should be large enough to make repetitions of events very rare. Combining the process samples gives the full
event dataset. Equivalently, one may draw a total count
$n\sim\Pois(\lambda(\boldsymbol{\theta}))$ and resample $n$ reference
indices independently with replacement using the mixture probabilities
\begin{equation}
  w_m(\boldsymbol{\theta})
  =\frac{\sum_s\lambda_s(\boldsymbol{\theta})
    w_m^{(s)}(\boldsymbol{\theta})}{\lambda(\boldsymbol{\theta})},
  \qquad \sum_m w_m(\boldsymbol{\theta})=1,
  \qquad \lambda(\boldsymbol{\theta})>0.
  \label{eq:total-event-weights}
\end{equation}
For cases with signed component coefficients, as in processes representing quantum interferences and higher-order perturbative corrections, 
the componentwise Poisson construction is unavailable. The total count
construction remains applicable provided the mixture probabilities in
\cref{eq:total-event-weights} are nonnegative.
The trained model and reference pool remain fixed across the ensemble.
These are unweighted pseudo-experiments from the finite empirical hNDE
model; increasing the reference-pool size improves its approximation to
the continuous surrogate.

Auxiliary observations are drawn from
$f(\cdot;\boldsymbol{\alpha})$ unless the construction explicitly
conditions on them. Repeated likelihood fits then give the sampling
distribution of $t_{\boldsymbol{\mu}}$ under the empirical hNDE model for
use in a Neyman construction~\cite{Neyman:1937}. These sampling capabilities address practical challenges encountered in the
ATLAS NSBI measurement of off-shell Higgs boson
production~\cite{ATLAS:2024tgo}. Inference with respect to the original simulator additionally requires independent validation or calibration against that simulator.

\subsection{Weighted unbinned Asimov inference}
\label{sec:asimov}
An unbinned Asimov dataset is a numerical quadrature for the expected
event measure. It is used extensively to study the expected power of the statistical analyses and their asymptotic behavior~\cite{Cowan:2010js}. In this subsection, we apply the finite weighted Asimov construction of
Ref.~\cite{LopesDeSa:finite-asimov} to hNDE. We then study the expectation
and variance of the log likelihood ratio and profiled test statistic, comparing fixed component normalizations with normalizations estimated on the same sample.

\subsubsection{Finite Asimov construction for hNDE}
\label{sec:asimov-construction}

At a generating point $\boldsymbol{\theta}_A=(\boldsymbol{\mu}_A,\boldsymbol{\alpha}_A)$,
write $h_A=h(\cdot;\boldsymbol{\theta}_A)$ and
$\lambda_A=\lambda(\boldsymbol{\theta}_A)$. The population event measure is
\begin{equation}
  \dd N_A(\mathbf{x})
  =\nu_A(\mathbf{x})\dd\mathbf{x}
  =\qRef(\mathbf{x})h_A(\mathbf{x})\dd\mathbf{x}.
  \label{eq:asimov-measure}
\end{equation}
Here $h$ uses the population-normalized ratios $r_{s,\boldsymbol{\psi}}$;
below, $\widetilde h$ uses their finite-sample normalized counterparts.

Keep the trained model fixed and choose reference points
$\{\mathbf{x}_m,\omega_m\}_{m=1}^{M}$ with
$\omega_m>0$ and $\sum_m\omega_m=1$.
For independent draws from $\qRef$, $\omega_m=1/M$.
Using the normalization of \cref{eq:empirical-ratio-normalization}, define
\begin{equation}
  Z_s(\boldsymbol{\theta})
  =\sum_m\omega_m r_{s,\boldsymbol{\psi}}
    (\mathbf{x}_m;\boldsymbol{\theta}),
  \qquad
  \widetilde h(\mathbf{x};\boldsymbol{\theta})
  =\sum_s\lambda_s(\boldsymbol{\theta})
    \frac{r_{s,\boldsymbol{\psi}}(\mathbf{x};\boldsymbol{\theta})}
         {Z_s(\boldsymbol{\theta})}.
  \label{eq:sample-dependent-normalizer}
\end{equation}
Here $\widetilde h$ approximates the population-normalized $h$ in
\cref{eq:h-definition}. For numerical integration and resampling, the
component probabilities on this fixed weighted reference sample are
\begin{equation}
  w_m^{(s)}(\boldsymbol{\theta})
  =\omega_m\widetilde r_{s,\boldsymbol{\psi}}
    (\mathbf{x}_m;\boldsymbol{\theta}),
  \qquad
  \widetilde r_{s,\boldsymbol{\psi}}
  =\frac{r_{s,\boldsymbol{\psi}}}{Z_s},
  \qquad \sum_m w_m^{(s)}(\boldsymbol{\theta})=1.
  \label{eq:component-weights}
\end{equation}
For independent direct-flow draws, $\omega_m=1/M$. These probabilities
are exactly normalized on the finite sample.

On this sample, the intensity-to-reference ratio $\widetilde h$ satisfies the exact identity
\begin{equation}
  \sum_m\omega_m\widetilde h(\mathbf{x}_m;\boldsymbol{\theta})
  =\lambda(\boldsymbol{\theta}).
  \label{eq:finite-intensity-normalization}
\end{equation}
Assigning the Asimov weights therefore gives
\begin{equation}
  w_m^A=\omega_m\widetilde h(\mathbf{x}_m;\boldsymbol{\theta}_A),
  \qquad
  \sum_m w_m^A=\lambda(\boldsymbol{\theta}_A).
  \label{eq:asimov-total-yield}
\end{equation}
The number of integration points $M$ controls the numerical approximation.
Their total weight is the expected event yield. The extended count term
is therefore represented independently of $M$. For an integrable function
$g_0$, the weighted events give the quadrature
\begin{equation}
  \int g_0(\mathbf{x})\dd N_A(\mathbf{x})
  \simeq\sum_m w_m^A g_0(\mathbf{x}_m).
  \label{eq:asimov-sample-expectation}
\end{equation}

Now choose auxiliary observations $\mathbf{a}_A$ such that
$f(\mathbf{a}_A;\boldsymbol{\alpha})$ has a positive, finite global maximum
at $\boldsymbol{\alpha}_A$. The finite weighted log likelihood is, up to
parameter-independent terms,
\begin{equation}
  \widehat\ell_A(\boldsymbol{\theta})
  =-\lambda(\boldsymbol{\theta})
   +\sum_m w_m^A\log\!\left[\qRef(\mathbf{x}_m)\widetilde h(\mathbf{x}_m;\boldsymbol{\theta})\right]
   +\log f(\mathbf{a}_A;\boldsymbol{\alpha}).
  \label{eq:asimov-loglikelihood}
\end{equation} 
The hat distinguishes this finite-sample objective
from its population counterpart. In a HEP likelihood, this is the direct
unbinned analogue of filling each histogram bin with its expected count
and setting the auxiliary global observables to their Asimov values.

The global-maximum property follows directly from
\cref{eq:finite-intensity-normalization}. As shown in Ref.~\cite{LopesDeSa:finite-asimov}, assuming that the normalizers and total
intensities at the reference points are finite and positive throughout the
parameter domain and writing
$u_m(\boldsymbol{\theta})=
\widetilde h(\mathbf{x}_m;\boldsymbol{\theta})/
\widetilde h(\mathbf{x}_m;\boldsymbol{\theta}_A)$ gives
\begin{equation}
  \widehat{\Delta\ell}(\boldsymbol{\theta})=\widehat\ell_A(\boldsymbol{\theta})
  -\widehat\ell_A(\boldsymbol{\theta}_A)
  ={}-\sum_m w_m^A
    \bigl[u_m(\boldsymbol{\theta})-1-\log u_m(\boldsymbol{\theta})\bigr]
  +\log\frac{f(\mathbf{a}_A;\boldsymbol{\alpha})}
                   {f(\mathbf{a}_A;\boldsymbol{\alpha}_A)}
  \leq0.
  \label{eq:finite-asimov-closure}
\end{equation}
Each bracket is nonnegative and the auxiliary term is nonpositive.
Thus $\boldsymbol{\theta}_A$ is a global maximizer, though the maximum need not be unique.

For differentiable models at an interior generating point, the exact score
cancellation follows by differentiating
\cref{eq:finite-intensity-normalization}:
\begin{align}
  \sum_m\omega_m\partial_a\widetilde h
       (\mathbf{x}_m;\boldsymbol{\theta})
  &=\partial_a\lambda(\boldsymbol{\theta}),\notag\\
  \left.\partial_a\widehat\ell_A\right|_{\boldsymbol{\theta}_A}
  &=-\left.\partial_a\lambda\right|_{\boldsymbol{\theta}_A}
    +\left.\sum_m\omega_m\partial_a\widetilde h
       (\mathbf{x}_m;\boldsymbol{\theta})\right|_{\boldsymbol{\theta}_A}
    +\left.\partial_a\log f(\mathbf a_A;\boldsymbol{\alpha})
       \right|_{\boldsymbol{\theta}_A}=0.
  \label{eq:finite-score-cancellation}
\end{align}
Here $\partial_a$ denotes differentiation with respect to $\theta_a$.
The global-maximum argument is stronger than this local score identity.

This result requires normalizing the complete parameter-dependent ratios,
including systematic shape factors, at every point evaluated during
profiling. The reference points and weights remain fixed within a scan.
When gradients are used, they must include the parameter dependence of
$Z_s$. Normalizing only nominal ratios generally does not ensure
closure in shape-dependent directions. This distinction matters in HEP
fits: signal strengths and pure normalization systematics can use fixed
component shapes, whereas detector or theory shape variations require the
complete interpolated ratios to be renormalized throughout profiling.
If only the nominal shapes are normalized, the exact guarantee remains
limited to yield-only directions.

Algorithm~\ref{alg:hNDE-direct-asimov} summarizes the construction.
The resulting profile statistic can use the generating point directly
as its unconditional maximum:
\begin{equation}
  \widehat t_A(\boldsymbol{\mu})
  =-2\left[
    \max_{\boldsymbol{\alpha}}
      \widehat\ell_A(\boldsymbol{\mu},\boldsymbol{\alpha})-\widehat\ell_A(\boldsymbol{\theta}_A)
    \right].
  \label{eq:finite-asimov-profile}
\end{equation}
It is nonnegative and vanishes at $\boldsymbol{\mu}_A$.
For a nonnegative scalar signal strength and $\mu_A>0$, the one-sided
discovery statistic is $q_{0,A}=\widehat t_A(0)$.
Under the usual large-sample and Wald approximations~\cite{Cowan:2010js},
\begin{equation}
  \sigma_A\simeq\frac{\mu_A}{\sqrt{q_{0,A}}},
  \qquad
  Z_A\simeq\sqrt{q_{0,A}},
  \label{eq:asimov-sigma}
\end{equation}
where $\sigma_A$ estimates the uncertainty on the signal strength and
$Z_A$ the median discovery significance.

Exact closure fixes the location of a likelihood maximum but the convergence of the scan shape and sensitivity still requires sufficient integration
accuracy. Increasing $M$ and comparing independent reference samples
assess this numerical error, while independent simulator-based validation
tests the learned model. Note that a flow is not required for finite-sample closure, but provides inexpensive, statistically independent reference points for the Asimov construction and
enables the importance-sampling application in \cref{sec:efficient-asimov}.
The following analysis quantifies the remaining integration error.
All training is held fixed; changing the quadrature does not retrain the
model or repair local classifier bias.

\begin{algorithm}[tbp!]
\setstretch{0.9}
\caption{hNDE Asimov construction}
\label{alg:hNDE-direct-asimov}
\begin{algorithmic}

\Require Frozen hNDE model
$\{\qRef,r_{s,\boldsymbol{\psi}},\lambda_s\}$;
generating point $\boldsymbol{\theta}_A$;
auxiliary likelihood $f$; sample size $M$

\Ensure Weighted Asimov sample $\mathcal A$, auxiliary observations
$\mathbf a_A$, and log-likelihood evaluator $\widehat\ell_A$
\vspace{0.5em}

\State Draw $\mathbf{x}_m\sim\qRef$ independently, set $\omega_m=1/M$,
and keep these points and weights fixed

\Function{$\widetilde h$}{$\mathbf{x}_m,\boldsymbol{\theta}$}
  \State Evaluate the complete component ratios and compute, for every $s$,
  \[
    Z_s(\boldsymbol{\theta})
    =\sum_m\omega_m r_{s,\boldsymbol{\psi}}
      (\mathbf{x}_m;\boldsymbol{\theta})
  \]
  \State \Return the vector with entries
  \[
    \widetilde h(\mathbf{x}_m,\boldsymbol{\theta})
    =\sum_s\lambda_s(\boldsymbol{\theta})
      \frac{r_{s,\boldsymbol{\psi}}(\mathbf{x}_m;\boldsymbol{\theta})}
           {Z_s(\boldsymbol{\theta})}
  \]
\EndFunction

\State Set $w_m^A=\omega_m\widetilde h(\mathbf{x}_m,\boldsymbol{\theta}_A)$ and
$\mathcal A=\{(\mathbf{x}_m,w_m^A)\}_{m=1}^{M}$

\State Verify $\sum_m w_m^A=\lambda(\boldsymbol{\theta}_A)$

\State Choose $\mathbf a_A$ such that
$\boldsymbol{\alpha}_A\in\arg\max_{\boldsymbol{\alpha}}
f(\mathbf a_A;\boldsymbol{\alpha})$

\State \Return $\mathcal A$, $\mathbf a_A$, and the function
\[
  \widehat\ell_A(\boldsymbol{\theta})
  =-\lambda(\boldsymbol{\theta})
   +\sum_mw_m^A\log\!\left[\qRef(\mathbf{x}_m)\widetilde h(\mathbf{x}_m,\boldsymbol{\theta})\right]
   +\log f(\mathbf a_A;\boldsymbol{\alpha}),
\]
calling $\widetilde h$ at every scan or profiling point

\end{algorithmic}
\end{algorithm}
\FloatBarrier

\subsubsection{Population log likelihood and expected test statistic}
\label{sec:asimov-population}

The population expectation and the expectation over numerical samples are
different operations. The former averages the event data generated at
$\boldsymbol{\theta}_A$, while keeping the auxiliary observations at
$\mathbf a_A$. The latter describes repeated numerical integrations of
that fixed learned model. 

Using the same convention as \cref{eq:asimov-loglikelihood}, the expected
extended log likelihood is
\begin{equation}
  \ell_A(\boldsymbol{\theta})
  =-\lambda(\boldsymbol{\theta})
   +\E_{\qRef}\!\left[
     h_A(\mathbf{x})\log \left(\qRef(\mathbf{x})h(\mathbf{x};\boldsymbol{\theta})\right)\right]
   +\log f(\mathbf a_A;\boldsymbol{\alpha}).
  \label{eq:expected-loglikelihood}
\end{equation}

For positive intensities and finite expectations, let
$u_{\boldsymbol{\theta}}=h(\cdot;\boldsymbol{\theta})/h_A
=\nu(\cdot;\boldsymbol{\theta})/\nu_A$. Then
\begin{align}
  \Delta\ell(\boldsymbol{\theta})
  &\equiv\ell_A(\boldsymbol{\theta})-\ell_A(\boldsymbol{\theta}_A)
    =-\int\nu_A(\mathbf{x})
       [u_{\boldsymbol{\theta}}-1-\log u_{\boldsymbol{\theta}}]
       \dd\mathbf{x}
    +\log\frac{f(\mathbf a_A;\boldsymbol{\alpha})}
                    {f(\mathbf a_A;\boldsymbol{\alpha}_A)}\leq0.
  \label{eq:population-asimov-closure}
\end{align}
The event term is the negative generalized Kullback--Leibler (KL) divergence
between the intensities. As in the finite sample case above, since the auxiliary likelihood is maximized at $\boldsymbol{\alpha}_A$, the generating point globally maximizes the full population objective. The event-level integrand and reference expectation for $-2\Delta\ell(\boldsymbol{\theta})$ are
\begin{equation}
  Y_{\boldsymbol{\theta}}(\mathbf{x})
  =h_A(\mathbf{x})\log
    \frac{h_A(\mathbf{x})}{h(\mathbf{x};\boldsymbol{\theta})},
  \qquad I_{\boldsymbol{\theta}}=\E_{\qRef}[Y_{\boldsymbol{\theta}}].
  \label{eq:Y-definition}
\end{equation}
Thus
\begin{align}
  -2\Delta\ell(\boldsymbol{\theta})
  ={}&2[\lambda(\boldsymbol{\theta})-\lambda_A
           +I_{\boldsymbol{\theta}}]-2\log\frac{f(\mathbf a_A;\boldsymbol{\alpha})}
                    {f(\mathbf a_A;\boldsymbol{\alpha}_A)}.
  \label{eq:asimov-loglikelihood-difference}
\end{align}
The common reference cancels inside the logarithm. Each reference event
contributes its log intensity ratio weighted by the generating intensity
relative to $\qRef$. The integrand $Y_{\boldsymbol{\theta}}$ can have either
sign even though $-2\Delta\ell(\boldsymbol{\theta})\geq 0$. The yield and auxiliary terms are evaluated
directly and contribute no direct quadrature error at fixed parameters.

The population Asimov test statistic is
\begin{equation}
  t_A(\boldsymbol{\mu})
  =\min_{\boldsymbol{\alpha}}\left[-2\Delta\ell(\boldsymbol{\mu},\boldsymbol{\alpha})\right].
  \label{eq:expectedteststatistic}
\end{equation}
Here the data expectation is taken before profiling. Consequently, $t_A$
is generally not the mean of the profile statistic over pseudo-experiments but its use for median expected sensitivity retains the asymptotic
assumptions in \cref{eq:asimov-sigma}.

\subsubsection{Expectation and variance with fixed and with same-sample normalization}
\label{sec:asimov-same-sample}

We first consider the case where the population-normalized ratios are used directly. For independent integration events $\mathbf{x}_m\sim\qRef$, replacing the population expectation by its sample mean gives
\begin{align}
  -2\widehat{\Delta\ell}^{\mathrm{fixed}}(\boldsymbol{\theta})
  &=2\left[\lambda(\boldsymbol{\theta})-\lambda_A
      +\frac1M\sum_m Y_{\boldsymbol{\theta}}(\mathbf{x}_m)\right]
    -2\log\frac{f(\mathbf a_A;\boldsymbol{\alpha})}
                     {f(\mathbf a_A;\boldsymbol{\alpha}_A)}.
  \label{eq:asimov-mc-estimator}
\end{align}
At a fixed tested point, these estimators are exactly unbiased over
numerical samples. When the second moments are finite,
\begin{align}
  \E_{\mathrm{MC}}[-2\widehat{\Delta\ell}^{\mathrm{fixed}}]
    &=-2\Delta\ell,
  &\Var_{\mathrm{MC}}[-2\widehat{\Delta\ell}^{\mathrm{fixed}}]
    &=\frac4M\Var_{\qRef}[Y_{\boldsymbol{\theta}}].
  \label{eq:fixed-normalization-moments}
\end{align}
All functions are evaluated at the same fixed $\boldsymbol{\theta}$.
The two log likelihoods in their difference use the same events, so their
covariance must be retained. Because of that the variance of
$\widehat{\Delta\ell}^{\mathrm{fixed}}$ is
$\Var_{\qRef}[Y_{\boldsymbol{\theta}}]/M$ and not the sum of two independent
log-likelihood variances.

A single normalization sample may instead be evaluated once and held fixed
while new integration samples are drawn. Conditional on that normalization
sample, the sample means remain unbiased for the resulting fixed integrands.
If its normalizers are inexact, however, the conditional target need not
equal the population objective in \cref{eq:expected-loglikelihood}.
Increasing only the integration sample size does not remove this fixed
normalization error. Also, fixed-normalization quadrature does not generally
place the finite-sample likelihood maximum exactly at the generating point.

In the finite construction, changing the integration sample changes both
the average and the normalizers inside its integrand. The additional
fluctuations are correlated with the direct sampling error. To propagate
them together, we use $\E_{\qRef}[r_{s,\boldsymbol{\psi}}]=1$ and the
finite-sample normalized ratios
$\widetilde r_{s,\boldsymbol{\psi}}=r_{s,\boldsymbol{\psi}}/Z_s$.
The trained functions, generating point, tested point, and auxiliary observations are fixed throughout this calculation.

Let $\delta Y_{\boldsymbol{\theta}}(\mathbf{x})$ denote the
first-order absolute change in the integrand
$Y_{\boldsymbol{\theta}}(\mathbf{x})$ caused by estimating the
component normalizers from this finite sample.
Since $I_{\boldsymbol{\theta}}=\E_{\qRef}[Y_{\boldsymbol{\theta}}]$,
the additional change in the integral is
$\E_{\qRef}[\delta Y_{\boldsymbol{\theta}}]$, where
\begin{equation*}
  \delta Y_{\boldsymbol{\theta}}
  =\left(\log\frac{h_A}{h_{\boldsymbol{\theta}}}+1\right)\delta h_A
   -\frac{h_A}{h_{\boldsymbol{\theta}}}\,
      \delta h_{\boldsymbol{\theta}}.
\end{equation*}
We abbreviate the intensity $h_{\boldsymbol{\theta}}:=h(\cdot;\boldsymbol{\theta})$, and $\delta h_A$ and
$\delta h_{\boldsymbol{\theta}}$ denote their first order absolute changes.
Thus, calculating the normalization contribution requires the changes
in both intensities.

To obtain these changes, abbreviate
$r_{s,A}=r_{s,\boldsymbol{\psi}}(\cdot;\boldsymbol{\theta}_A)$ and
$r_{s,\boldsymbol{\theta}}=r_{s,\boldsymbol{\psi}}(\cdot;\boldsymbol{\theta})$,
and similarly for the yields and normalizers.
For either point, the normalizer fluctuation about unity is
\begin{equation*}
  \delta Z_{s,v}
  =Z_{s,v}-1
  =\frac1M\sum_m r_{s,v}(\mathbf{x}_m)-1,
  \qquad v=A,\boldsymbol{\theta}.
\end{equation*}
The finite-sample normalized ratio is
$r_{s,v}/(1+\delta Z_{s,v})$. Expanding
$1/(1+\delta Z)\simeq1-\delta Z$ therefore gives
\begin{equation*}
  \delta h_v
  =-\sum_s\lambda_{s,v}r_{s,v}\,\delta Z_{s,v}.
\end{equation*}
Substituting these intensity changes into
$\delta Y_{\boldsymbol{\theta}}$ and averaging over $\qRef$ gives
\begin{equation*}
  \E_{\qRef}[\delta Y_{\boldsymbol{\theta}}]
  =-\sum_s b^A_{s,\boldsymbol{\theta}}\,\delta Z_{s,A}
   +\sum_s b^T_{s,\boldsymbol{\theta}}\,
      \delta Z_{s,\boldsymbol{\theta}},
\end{equation*}
where
\begin{align}
  b^A_{s,\boldsymbol{\theta}}
  &=\lambda_s(\boldsymbol{\theta}_A)
    \E_{\qRef}\!\left[
      r_{s,A}\left(\log\frac{h_A}{h_{\boldsymbol{\theta}}}+1\right)
    \right],\notag\\
  b^T_{s,\boldsymbol{\theta}}
  &=\lambda_s(\boldsymbol{\theta})
    \E_{\qRef}\!\left[
      r_{s,\boldsymbol{\theta}}\frac{h_A}{h_{\boldsymbol{\theta}}}
    \right].
  \label{eq:normalization-coefficients}
\end{align}
Here the expectation is over $\mathbf{x}$, with the normalizer errors
from the given sample held fixed. The coefficients quantify the
sensitivity of the integral to those errors ($A$ and $T$ refer to
the generating and tested intensities).

Each $\delta Z$ is itself a sample average of $r-1$. Substituting
these averages into the normalization correction allows both sources
of error to be written as one event average, identifying
\begin{align}
  U_{\boldsymbol{\theta}}(\mathbf{x})
  ={}&Y_{\boldsymbol{\theta}}(\mathbf{x})-I_{\boldsymbol{\theta}}
    -\sum_s b^A_{s,\boldsymbol{\theta}}
       [r_{s,A}(\mathbf{x})-1]+\sum_s b^T_{s,\boldsymbol{\theta}}
       [r_{s,\boldsymbol{\theta}}(\mathbf{x})-1],
  \qquad \E_{\qRef}[U_{\boldsymbol{\theta}}]=0.
  \label{eq:normalization-influence}
\end{align}
The difference $Y_{\boldsymbol{\theta}}-I_{\boldsymbol{\theta}}$
accounts for ordinary sampling fluctuations, while the two sums
account for the normalization changes. Combining them in
$U_{\boldsymbol{\theta}}$ retains the correlations caused by using
the same events for both calculations.

We assume positive finite sample normalizers and finite second moments for $Y_{\boldsymbol{\theta}}$ and $r_{s,v}$. The delta method then gives
\begin{align}
  -2\widehat{\Delta\ell}(\boldsymbol{\theta})
     +2\Delta\ell(\boldsymbol{\theta})
  =\frac2M\sum_m U_{\boldsymbol{\theta}}(\mathbf{x}_m)
     +o_p(M^{-1/2}).
  \label{eq:asimov-mc-expansion}
\end{align}
Here $o_p(M^{-1/2})$ is negligible in probability at that scale. To infer
variances, the remainders must also be negligible at that scale in mean
square. For the bias orders below, assume an integrable second-order Taylor
remainder with expectation of order $M^{-1}$, then
\begin{align}
  \E_{\mathrm{MC}}[-2\widehat{\Delta\ell}(\boldsymbol{\theta})]
    &=-2\Delta\ell(\boldsymbol{\theta})+O(M^{-1}),
  \label{eq:same-sample-expectations}\\
  \Var_{\mathrm{MC}}[-2\widehat{\Delta\ell}(\boldsymbol{\theta})]
    &=\frac4M\Var_{\qRef}[U_{\boldsymbol{\theta}}]+o(M^{-1}).
  \label{eq:asimov-mc-variance}
\end{align}
Unlike fixed-normalization sample means, the same-sample
estimators are nonlinear functions of empirical averages and need not be
exactly unbiased. The additional terms in $U$ change the leading
variance coefficient. Their covariances can increase or decrease it, so
normalization is not simply an independent positive variance contribution.
At $\boldsymbol{\theta}=\boldsymbol{\theta}_A$, $U=0$ identically and
$-2\widehat{\Delta\ell}=0$ exactly for every sample. 

For HEP signal-strength measurements with fixed component shapes, including
pure yield systematics, $r_{s,A}=r_{s,\boldsymbol{\theta}}=r_s$.
The leading difference fluctuation simplifies to
\begin{equation}
  U_{\boldsymbol{\theta}}
  =Y_{\boldsymbol{\theta}}-I_{\boldsymbol{\theta}}
   -\sum_s(b^A_{s,\boldsymbol{\theta}}-b^T_{s,\boldsymbol{\theta}})
          (r_s-1).
  \label{eq:yield-only-influence}
\end{equation}
The normalizers remain random across independently drawn samples even
though they do not vary along yield-only parameter directions. Detector
and theory shape systematics require the full generating- and tested-point
terms in \cref{eq:normalization-influence}.

For comparison, suppose a separate, independent sample of size $N$ is used
to estimate all normalizers, while $M$ events estimate the likelihood
integral. Define the normalization-only contribution
$D_{\boldsymbol{\theta}}=U_{\boldsymbol{\theta}}
-(Y_{\boldsymbol{\theta}}-I_{\boldsymbol{\theta}})$.
For independent direct-flow samples and the same moment controls,
\begin{equation}
  \Var_{\mathrm{MC}}[-2\widehat{\Delta\ell}^{\mathrm{separate}}]
  =4\left\{\frac{\Var_{\qRef}[Y_{\boldsymbol{\theta}}]}{M}
          +\frac{\Var_{\qRef}[D_{\boldsymbol{\theta}}]}{N}\right\}
     +o(M^{-1}+N^{-1}).
  \label{eq:separate-normalization-variance}
\end{equation}
Its pointwise bias is $O(N^{-1})$ under a second-order expansion. These independent-sample variances add because the two samples are independent. Using the same sample instead requires their covariance and gives \cref{eq:asimov-mc-variance}.

\subsubsection{Expectation and variance after profiling}
\label{sec:asimov-profiling}

For a fixed POI value, define the population conditional optimum by
\begin{equation}
  \boldsymbol{\alpha}^{*}_{\boldsymbol{\mu}}
  =\arg\min_{\boldsymbol{\alpha}}\left[-2\Delta\ell(\boldsymbol{\mu},\boldsymbol{\alpha})\right],
  \qquad
  \boldsymbol{\theta}^{*}_{\boldsymbol{\mu}}
  =(\boldsymbol{\mu},\boldsymbol{\alpha}^{*}_{\boldsymbol{\mu}}),
  \label{eq:population-profile-optimizer}
\end{equation}
so that $t_A(\boldsymbol{\mu})=-2\Delta\ell(\boldsymbol{\theta}^{*}_{\boldsymbol{\mu}})$.
Suppose this minimizer is unique and interior, its NP Hessian
$H_{\boldsymbol{\mu}}$ is positive definite, and the finite-sample
conditional minimizer $\widehat{\widehat{\boldsymbol{\alpha}}}_{\boldsymbol{\mu},A}$
is consistent. Assume uniform smooth stochastic expansions for
the objective and its gradient near this optimum. Writing
$g_M=-2\nabla_{\boldsymbol{\alpha}}(\widehat{\Delta\ell}-\Delta\ell)
 (\boldsymbol{\theta}^{*}_{\boldsymbol{\mu}})=O_p(M^{-1/2})$, the
sample-average optimization expansion~\cite{Shapiro2000MonteCarlo} gives
\begin{align}
  \widehat{\widehat{\boldsymbol{\alpha}}}_{\boldsymbol{\mu},A}
       -\boldsymbol{\alpha}^{*}_{\boldsymbol{\mu}}
  &=-H_{\boldsymbol{\mu}}^{-1}g_M+o_p(M^{-1/2}),\notag\\
  \widehat t_A(\boldsymbol{\mu})
  &=-2\widehat{\Delta\ell}(\boldsymbol{\theta}^{*}_{\boldsymbol{\mu}})
    -\tfrac12 g_M^{\mathsf T}H_{\boldsymbol{\mu}}^{-1}g_M
    +o_p(M^{-1}).
  \label{eq:profile-optimization-correction}
\end{align}
The fitted NPs move at order $M^{-1/2}$, but their extra
effect on the objective is second order because the population gradient
vanishes. For same-sample normalization, the unconditional maximum is
already fixed at the generating point by exact closure. Consequently,
\begin{equation}
  \widehat t_A(\boldsymbol{\mu})-t_A(\boldsymbol{\mu})
  =\frac2M\sum_m U_{\boldsymbol{\theta}^{*}_{\boldsymbol{\mu}}}
       (\mathbf{x}_m)+o_p(M^{-1/2}).
  \label{eq:profiled-asimov-expansion}
\end{equation}
With the expectation and mean-square controls stated above,
\begin{align}
  \E_{\mathrm{MC}}[\widehat t_A(\boldsymbol{\mu})]
    &=t_A(\boldsymbol{\mu})+O(M^{-1}),\notag\\
  \Var_{\mathrm{MC}}[\widehat t_A(\boldsymbol{\mu})]
    &=\frac4M\Var_{\qRef}
       [U_{\boldsymbol{\theta}^{*}_{\boldsymbol{\mu}}}]+o(M^{-1}).
  \label{eq:profiled-asimov-moments}
\end{align}
Thus profiling contributes only $o(M^{-1})$ to the variance relative to
evaluation at the population conditional optimum. It can contribute an
$O(M^{-1})$ bias in addition to the normalization bias.

Without finite-sample renormalization, the same leading results hold with
$U$ replaced by $Y-I$. The finite-sample unconditional maximum may also
move. Its optimization contributes another $O_p(M^{-1})$ term under the
corresponding regularity conditions. Pointwise unbiasedness in
\cref{eq:fixed-normalization-moments} therefore does not imply an exactly
unbiased profiled statistic.

\subsection{Efficient Asimov samples}
\label{sec:efficient-asimov}
The role of the reference flow extends beyond generating an arbitrarily
large independent sample. It also supplies an explicit map from a simple
latent distribution to the event observables. For a standard Gaussian base,
\begin{equation}
  \mathbf{x}=T_{\boldsymbol{\phi}}(\mathbf{z}),
  \qquad \mathbf{z}\sim p_0=\Normal(0,\mathbf I),
  \label{eq:flow-transport}
\end{equation}
where transporting $p_0$ through $T_{\boldsymbol{\phi}}$ gives $\qRef$.
The integral entering the Asimov likelihood ratio can therefore be written
\begin{equation}
  I_{\boldsymbol{\theta}}
  =\E_{\qRef}[Y_{\boldsymbol{\theta}}(\mathbf{x})]
  =\E_{p_0}\!\left[
    Y_{\boldsymbol{\theta}}(T_{\boldsymbol{\phi}}(\mathbf z))\right].
  \label{eq:latent-asimov-integral}
\end{equation}
Knowing this map allows us to choose how the latent space is sampled.
Two complementary improvements are possible: latent points can cover the
base distribution more uniformly, or the sampling density can concentrate
evaluations where they most reduce the integration error.

Randomized quasi-Monte Carlo (RQMC) is a natural option for the first approach. A scrambled
Sobol sequence $\mathbf u_m\in(0,1)^d$ is transported through the
componentwise Gaussian quantile and the flow:
\begin{equation}
  \mathbf u_m\longmapsto\mathbf z_m=\Phi^{-1}(\mathbf u_m)
  \longmapsto\mathbf x_m=T_{\boldsymbol{\phi}}(\mathbf z_m).
  \label{eq:rqmc-transport}
\end{equation}
Low-discrepancy points cover the latent hypercube more uniformly than
independent points. For sufficiently regular transformed integrands, this
can reduce integration error and give convergence faster than the usual
$M^{-1/2}$ MC rate~\cite{owen1997scrambled,andral2024combining,klebanov2023transporting}.
The relevant integrands include those used to estimate the component
normalizers.

Neural importance sampling is a natural option for the second approach. The evaluable reference density lets us account for the changed proposal through importance weights and reduce the integration sample size needed for a given precision, while keeping the trained hNDE model fixed. Note the two approaches can be used together, but in this paper we will demonstrate the NIS approach explicitly and reserve a joint approach for future work.

\subsubsection{Neural importance sampling}
\label{sec:nis-proposal}

The statistical analysis in \cref{sec:asimov} identifies the event
contribution $U_{\boldsymbol{\theta}}$, including same-sample normalization, to the variance of the Asimov test statistic. As demonstrated in App.~\ref{app:importance-normalization}, for an alternative proposal density $g(\mathbf{x})$ whose support covers $\qRef$, the same Asimov integral is
\begin{equation}
  I_{\boldsymbol{\theta}}
  =\E_g\!\left[
  \frac{\qRef(\mathbf{x})}{g(\mathbf{x})}Y_{\boldsymbol{\theta}}(\mathbf{x})
  \right].
  \label{eq:asimov-importance}
\end{equation}
For the same-sample normalization, the leading event fluctuation under importance sampling is $(\qRef/g)U_{\boldsymbol{\theta}}$. Its expectation is zero, including when the reference weights are self-normalized as in \cref{eq:self-normalized-weights}. Thus the leading variance of the finite Asimov statistic is $4/M$ times
\begin{equation}
  \E_g\!\left[
    \left(\frac{\qRef(\mathbf{x})}{g(\mathbf{x})}
    U_{\boldsymbol{\theta}}(\mathbf{x})\right)^2
  \right]
  =\int\frac{q_{\boldsymbol{\phi}}^2(\mathbf{x})
    U_{\boldsymbol{\theta}}^2(\mathbf{x})}{g(\mathbf{x})}\dd\mathbf{x}.
  \label{eq:importance-second-moment}
\end{equation}

As demonstrated in App.~\ref{app:nis-proposals}, for a single tested point, minimizing the variance under the alternative reference density $g$ gives the formal
optimum
\begin{equation}
  g_{\boldsymbol{\theta}}^\star(\mathbf{x})
  =\frac{\qRef(\mathbf{x})|U_{\boldsymbol{\theta}}(\mathbf{x})|}
         {\int\qRef(\mathbf{x}')|U_{\boldsymbol{\theta}}(\mathbf{x}')|
           \dd\mathbf{x}'},
  \label{eq:optimal-point-proposal}
\end{equation}
provided the denominator is finite and nonzero. Relative to sampling
from $\qRef$, this proposal increases the sampling frequency where
the absolute error contribution is large. 

For a complete scan, one proposal must provide useful precision at
several parameter points. Discovery studies may emphasize the null hypothesis, usually $\mu=0$, while exclusion studies may prioritize larger signal strengths.
Confidence-interval calculations require accuracy over the range containing
the interval crossings.
For design points $\{\boldsymbol{\theta}_k\}$ with fixed weights
$a_k\geq0$, we minimize the weighted sum of their individual variances,
or equivalently their second moments:
\begin{equation}
  \sum_k a_k\E_g\!\left[
    \left(\frac{\qRef}{g}U_{\boldsymbol{\theta}_k}\right)^2\right]
  =\int\frac{\qRef^2(\mathbf{x})}{g(\mathbf{x})}
    \sum_k a_kU_{\boldsymbol{\theta}_k}^2(\mathbf{x})\dd\mathbf{x}.
  \label{eq:scan-second-moment}
\end{equation}
The same minimization gives
\begin{equation}
  g^\star(\mathbf{x})
  =\frac{\qRef(\mathbf{x})A(\mathbf{x})}{\E_{\qRef}[A]},
  \qquad
  A(\mathbf{x})
  =\left[\sum_k a_kU_{\boldsymbol{\theta}_k}^2(\mathbf{x})\right]^{1/2},
  \label{eq:optimal-scan-proposal}
\end{equation}
provided $0<\E_{\qRef}[A]<\infty$. The function $A$ combines the size
of an event's error contributions across the chosen scan points.
The weights $a_k$ specify their relative priority, allowing the
proposal to emphasize particular hypotheses or anticipated
confidence-interval crossings.
The simpler target $A=|Y_{\boldsymbol{\theta}}|$, or its root-mean-square (RMS)
counterpart across a scan, avoids estimating the normalization
sensitivities. We also find this choice effective empirically, although
it does not account for the full normalization contribution to the
integration error.

Following the NIS approach~\cite{muller2019neuralimportance}, we approximate the proposal with a second flow model. A separate pilot
sample first identifies which regions contribute most to the error.
We draw $P$ events $\mathbf{x}^{\mathrm{pilot}}_p\sim\qRef$ and evaluate
\(\widehat U_{pk}
  :=\widehat U_{\boldsymbol{\theta}_k}
    (\mathbf{x}^{\mathrm{pilot}}_p)\) at each point.
Equation~\ref{eq:optimal-scan-proposal} then gives
the training weights $\widehat A_p$. These weights emphasize pilot
events with large error contributions across the scan. We train the
proposal flow $g_{\boldsymbol{\eta}}$ by maximizing
\begin{equation}
  \sum_{p=1}^{P}\widehat A_p
    \log g_{\boldsymbol{\eta}}(\mathbf{x}^{\mathrm{pilot}}_p).
  \label{eq:nis-weighted-training}
\end{equation}
After training, the proposal is held fixed and a fresh Asimov sample,
independent of the pilot, is drawn for integration.  The resulting sample can be reused throughout the likelihood scan, with the component normalizers recomputed at every evaluated
parameter point. The gain in precision comes from allocating events
more efficiently while preserving the integral of the frozen hNDE
model. The NIS construction presented here generalizes to any normalized NDE model, as demonstrated in App.~\ref{app:nde-importance}.

\subsubsection{Defensive importance sampling}
\label{sec:defensive-importance}
The formal proposal in \cref{eq:optimal-scan-proposal} can vanish where
$A$ vanishes, and the learned proposal can underestimate relevant tails.
To retain the full reference support, we use a defensive
mixture~\cite{Hesterberg01051995},
\begin{equation}
  g_\epsilon(\mathbf{x})
  =(1-\epsilon)g_{\boldsymbol{\eta}}(\mathbf{x})
   +\epsilon\qRef(\mathbf{x}),
  \qquad 0<\epsilon<1.
  \label{eq:defensive-mixture}
\end{equation}
Each event is drawn from $\qRef$ with probability $\epsilon$ and from
$g_{\boldsymbol{\eta}}$ otherwise. The importance weight uses the full
mixture density, regardless of which component generated the event
\begin{equation}
  \rho(\mathbf{x})
  =\frac{\qRef(\mathbf{x})}{g_\epsilon(\mathbf{x})}
  \leq\frac1\epsilon.
  \label{eq:raw-importance-weight}
\end{equation}
Thus $\epsilon$ controls the fraction of events drawn from the
reference and sets an upper bound on the raw importance weight. To create sample events for an Asimov sample, we hold the trained proposal, including $\epsilon$, fixed and draw a fresh sample of $M$ independent events
$\mathbf{x}_m\sim g_\epsilon$.
The self-normalized reference weights are
\begin{equation}
  \omega_m=\frac{\rho_m}{\sum_{\ell=1}^{M}\rho_\ell},
  \qquad \rho_m=\rho(\mathbf{x}_m),
  \qquad \sum_m\omega_m=1.
  \label{eq:self-normalized-weights}
\end{equation}
Weighted averages with these weights approximate expectations under
$\qRef$. Use the same weights throughout the Asimov construction. In particular, the finite-sample normalization becomes
\begin{equation}
  Z_s(\boldsymbol{\theta})
  =\sum_m\omega_m r_{s,\boldsymbol{\psi}}
    (\mathbf{x}_m;\boldsymbol{\theta}),
  \qquad
  \widetilde r_{s,\boldsymbol{\psi}}(\mathbf{x};\boldsymbol{\theta})
  =\frac{r_{s,\boldsymbol{\psi}}(\mathbf{x};\boldsymbol{\theta})}
         {Z_s(\boldsymbol{\theta})}.
  \label{eq:nis-ratio-normalization}
\end{equation}
The resulting Asimov weights are
$w_m^A=\omega_m\widetilde h(\mathbf{x}_m;\boldsymbol{\theta}_A)$,
with $\widetilde h=\sum_s\lambda_s\widetilde r_{s,\boldsymbol{\psi}}$.
They satisfy $\sum_m w_m^A=\lambda(\boldsymbol{\theta}_A)$ exactly.

Algorithm~\ref{alg:hNDE-nis-asimov} summarizes the procedure. The final weighted reference sample
replaces the events drawn from the reference flow in Algorithm~\ref{alg:hNDE-direct-asimov};
the remaining Asimov construction is unchanged. The design points guide
proposal training, while the resulting likelihood can be evaluated and
profiled throughout the parameter region of interest.

\begin{algorithm}[tbp!]
\setstretch{0.9}
\caption{hNDE Asimov construction with NIS}
\label{alg:hNDE-nis-asimov}
\begin{algorithmic}

\Require Frozen hNDE model $\{\qRef,r_{s,\boldsymbol{\psi}},\lambda_s\}$;
generating point $\boldsymbol{\theta}_A$; auxiliary likelihood $f$;
scan design $\{\boldsymbol{\theta}_k,a_k\}_{k=1}^{K}$;
pilot size $P$; integration sample size $M$;
defensive fraction $0<\epsilon<1$

\Ensure Weighted Asimov sample $\mathcal A$, auxiliary observations
$\mathbf a_A$, and log-likelihood evaluator $\widehat\ell_A$
\vspace{0.5em}

\Statex \textit{Phase 1: proposal training}

\State Draw $\mathbf{x}^{\mathrm{pilot}}_p\sim\qRef$,
$p=1,\ldots,P$, independently

\State Estimate $\widehat U_{pk}=\widehat U_{\boldsymbol{\theta}_k}(\mathbf{x}^{\mathrm{pilot}}_p)$ on the pilot sample using
\cref{eq:normalization-influence}

\State Set the training weights
\[
  \widehat A_p
  =\left[\sum_{k=1}^{K}a_k\widehat U_{pk}^{\,2}\right]^{1/2}
\]

\State Fit $g_{\boldsymbol{\eta}}$ with training weights $\widehat A_p$
using \cref{eq:nis-weighted-training}

\Statex \textit{Phase 2: importance sampling}

\State Form and freeze the proposal
\[
  g_\epsilon=(1-\epsilon)g_{\boldsymbol{\eta}}+\epsilon\qRef
\]

\State Draw $M$ fresh, independent events
$\mathbf{x}_m\sim g_\epsilon$

\State Compute the reference weights
\[
  \rho_m=\frac{\qRef(\mathbf{x}_m)}{g_\epsilon(\mathbf{x}_m)},
  \qquad
  \omega_m=\frac{\rho_m}{\sum_{\ell=1}^{M}\rho_\ell}
\]

\Statex \textit{Phase 3: Asimov construction}

\State Use the fixed sample $\{\mathbf{x}_m,\omega_m\}$ in
Algorithm~\ref{alg:hNDE-direct-asimov}, starting from its normalization step
\State \Return $\mathcal A$, $\mathbf a_A$, and $\widehat\ell_A$

\end{algorithmic}
\end{algorithm}

\section{Demonstration using toy model}
\label{sec:demonstration}
We use an example where the statistical model consists of a signal and a background, with each observation spanning five dimensions, to test the
three algorithms.
The example is inspired by several features of experimental HEP inference: a rare signal
within a much larger background, correlated observables affected by
detector response, classifier preselection, and an extended likelihood
with a signal strength and a shape NP. The methods, however, are of broad application.
Analytic reconstructed densities in this toy example provide a benchmark for the learned
model. The tests address four questions: whether hNDE reproduces the
target distributions and likelihood, whether the finite Asimov sample
returns its generating point, how much NIS improves integration
precision, and whether closure is retained when an NP modifying the high-dimensional distribution is profiled.

\subsection{Statistical model and training samples}
\label{sec:demonstration-model}

Signal and background each follow a correlated Gaussian mixture in
five latent variables, with two process-specific modes and a common
broad component. Gaussian detector smearing produces the reconstructed
observables $\mathbf{x}=(x_1,\ldots,x_5)$. This construction provides
overlapping signal and background distributions and analytically
evaluable reconstructed densities.

The inclusive expected yields are $\lambda_B^{\mathrm{incl}}=10^6$
and $\lambda_S^{\mathrm{incl}}=1.1\times10^3$. A classifier
preselection is chosen to give an expected background-to-signal ratio
of approximately $250$. The selected yields are
\begin{equation}
  \lambda_S=611.60,
  \qquad
  \lambda_B=152{,}822.48.
  \label{eq:postselection-yields}
\end{equation}
The yields emulate rare-signal searches in HEP. Preselection enriches
the signal while keeping a background-dominated region manageable for
this study.

All densities below refer to the selected region and are normalized
within it. For the nominal measurement, the signal strength $\mu\geq0$
is the only parameter. The statistical model is
\begin{align}
  &\nu(\mathbf{x};\mu)
  =\sum_{s}\lambda_s(\mu)p_s(\mathbf{x})
    =\mu\,\lambda_Sp_S(\mathbf{x})+\lambda_Bp_B(\mathbf{x}),\qquad
  &\lambda(\mu)=\mu\lambda_S+\lambda_B.
  \label{eq:demonstration-statistical-model}
\end{align}
Here $\lambda_S$ denotes the signal yield at $\mu=1$ and the background
yield is fixed. Inference uses the extended likelihood of
\cref{eq:extended-likelihood}. The final test adds a shape NP to the component densities.

Following Algorithm~\ref{alg:hNDE}, we train the reference flow on
$50{,}000$ selected signal events and $50{,}000$ selected background
events with equal component normalization. Separate classifier
ensembles then estimate $p_S/\qRef$ and $p_B/\qRef$, using five million
events per class: selected target events and fresh flow-generated
reference events. Each ratio is the arithmetic mean of four network
estimates. The baseline calculations use a fixed five-million-event
reference sample for empirical normalization. In the sample-size
comparisons below, the ratios are normalized again on each integration
sample, as required by the Asimov construction. Network architectures
and training settings are given in App.~\ref{app:demonstration-training}.
\subsection{Accuracy of the hybrid model}
\label{sec:demonstration-accuracy}

This test evaluates the density and likelihood accuracy of
Algorithm~\ref{alg:hNDE}. On a held-out validation partition, we check
classifier calibration, reweighting of the reference to each process,
and agreement of the reconstructed densities with analytic truth.
Before empirical normalization, the mean signal and background ratios
on the reference sample are $0.9992$ and $0.9998$, respectively.
The background projections and pairwise correlations are shown in
\cref{fig:hybrid-closure} in the appendix.

\Cref{fig:calibration} compares the reconstructed signal log density
with analytic truth. Across the full validation sample, the
correlations between reconstructed and analytic log densities are
$0.981$ for signal and $0.985$ for background. These checks test the
absolute densities obtained by multiplying the learned ratios by the
reference density. Their adequacy for inference is assessed directly
with a likelihood comparison.

\begin{figure}[htbp!]
  \centering
  \includegraphics[width=0.5\linewidth]
    {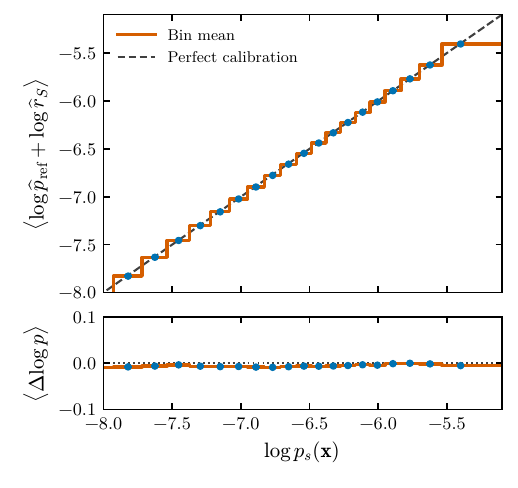}
  \caption{Signal-density validation against analytic truth after
  preselection. Events are grouped into equal-population bins of
  $\log p_S(\mathbf{x})$. The upper panel compares the mean reconstructed
  log density, $\log\qRef+\log\widetilde r_{S,\boldsymbol{\psi}}$,
  with the identity line. The lower panel shows the mean difference
  between reconstructed and analytic log densities. Vertical bars
  give the standard error of each bin mean.}
  \label{fig:calibration}
\end{figure}

\begin{figure}[htbp!]
  \centering
  \begin{subfigure}[t]{0.49\textwidth}
    \centering
    \includegraphics[width=\textwidth]
      {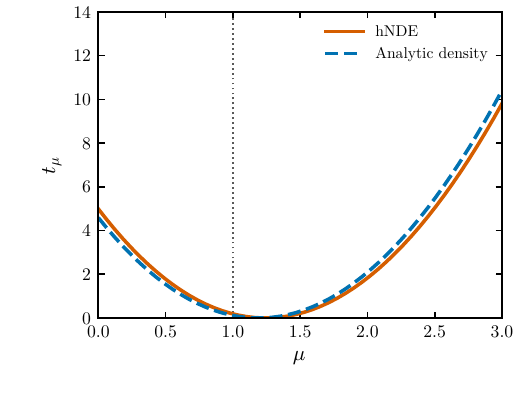}
    \caption{Validation against analytic truth}
    \label{fig:profile-hybrid-truth}
  \end{subfigure}
  \hfill
  \begin{subfigure}[t]{0.49\textwidth}
    \centering
    \includegraphics[width=\textwidth]
      {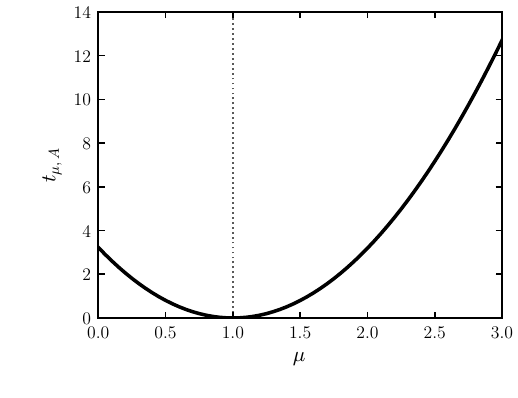}
    \caption{Asimov benchmark}
    \label{fig:asimov-benchmark}
  \end{subfigure}
  \caption{Likelihood validation and numerical benchmark.
  Panel~(\protect\subref{fig:profile-hybrid-truth}) compares the hNDE
  and analytic likelihood ratio scans on the same finite weighted
  simulation sample. Panel~(\protect\subref{fig:asimov-benchmark})
  shows the independent one million event direct-reference Asimov
  scan used to benchmark NIS. The vertical lines mark the generating
  signal strength $\mu=1$.}
  \label{fig:profile-asimov-comparison}
\end{figure}

We fit the hNDE and analytic models to the same independent finite
weighted validation sample. The fitted signal strengths are
$\widehat\mu=1.24$ and $1.19$, respectively. Their difference of
$0.05$ is small compared with the expected statistical uncertainty
of approximately $0.56$. The corresponding scans are shown in
\cref{fig:profile-hybrid-truth}. This comparison isolates the effect
of the learned model on inference for that validation sample. For precision analyses, agreement of the learned densities must
translate into sufficiently small shifts of fitted parameters and
likelihood-ratio scans. The small shift here illustrates the relevant
validation criterion.

For comparison, \cref{fig:flow-only-profile-comparison} in the
appendix shows the likelihood scan obtained by modeling the signal
and background densities directly with separate ensembles of
normalizing flows. These models use the same training data as the
hNDE classifiers. Their scan reaches its minimum at
$\widehat\mu\simeq3$, substantially displaced from the analytic
minimum on the same validation sample. The hybrid construction
achieves much closer agreement with the analytic likelihood in
this example, illustrating the value of hNDE over traditional flow-based density estimation.

\subsection{Asimov closure and expected sensitivity}
\label{sec:demonstration-asimov}

This test checks the finite sample closure of
Algorithm~\ref{alg:hNDE-direct-asimov} and the sensitivity predictions
obtained from the resulting likelihood. We construct the Asimov sample
at $\mu_A=1$ from the five-million-event reference sample. Its total
weight is $\lambda_S+\lambda_B$, or approximately $153{,}434$.
The likelihood score at the generating point is
$6.8\times10^{-13}$ and the fit returns $\widehat\mu_A=1.00$,
confirming the expected internal closure. The five-million-event integration sample is of the same order as the million-event samples discussed for the ATLAS NSBI implementation~\cite{ATLAS:2024rpr}.

The discovery statistic, expected significance, and Asimov uncertainty
are
\begin{equation}
  q_{0,A}=3.230,
  \qquad Z_A=1.797,
  \qquad \sigma_A=0.556.
  \label{eq:exercise5-asimov}
\end{equation}
The uncertainty estimated from the local likelihood curvature is
$0.558$. An independent Asimov calculation using one million events drawn
from the reference flow gives $q_{0,A}=3.227\pm0.003$, with the
uncertainty estimated using ten blocks.
This agrees with the larger-sample result and supplies the numerical
benchmark for the NIS comparison in
\cref{sec:demonstration-nis,fig:asimov-benchmark}.

\begin{figure}[htbp!]
  \centering
  \begin{subfigure}[t]{0.48\textwidth}
    \centering
    \includegraphics[width=\linewidth]
      {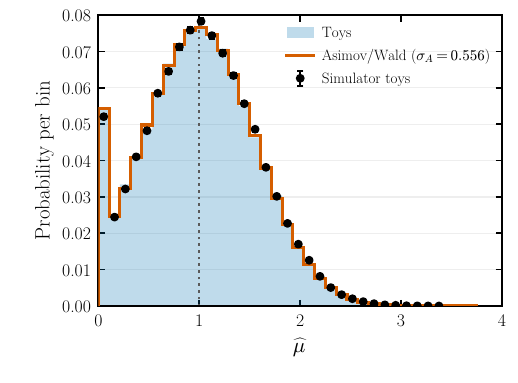}
    \caption{Signal-strength estimator.}
    \label{fig:asimov-toys-estimator}
  \end{subfigure}
  \hfill
  \begin{subfigure}[t]{0.48\textwidth}
    \centering
    \includegraphics[width=\linewidth]
      {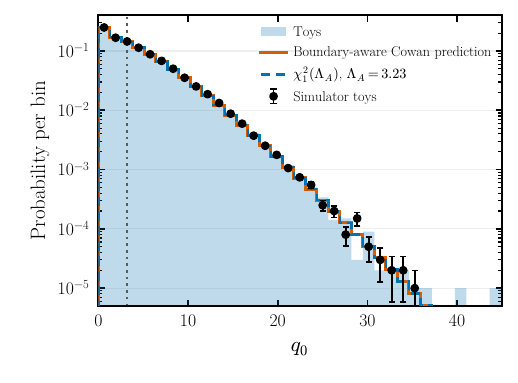}
    \caption{Discovery test statistic.}
    \label{fig:asimov-toys-discovery}
  \end{subfigure}
  \caption{Combined Asimov and simulator validation at
  $\mu_{\mathrm{true}}=1$, with $100{,}000$ pseudo-experiments in each
  ensemble. The filled blue histograms (``Toys'') show hNDE-generated
  pseudo-experiments; the black markers (``Simulator toys'') show
  simulator-generated pseudo-experiments with Poisson statistical error
  bars. Both ensembles are analyzed with the same frozen hybrid likelihood.
  Panel~(\protect\subref{fig:asimov-toys-estimator}) compares the
  maximum-likelihood estimator distribution with the Wald prediction
  based on $\sigma_A=0.556$.
  Panel~(\protect\subref{fig:asimov-toys-discovery}) compares the
  discovery statistic $q_0$ on a logarithmic vertical scale with the boundary-aware Asimov prediction
  and the noncentral-$\chi^2$ approximation.}
  \label{fig:asimov-prediction-vs-toys}
\end{figure}

We also compare the Asimov predictions with $100{,}000$
pseudo-experiments generated from the frozen hNDE model at $\mu=1$
after preselection. The fitted signal strength has mean $1.000$ and standard deviation
$0.54$, compared with the Asimov uncertainty $0.56$.
The fraction of fits at the physical boundary is $3.64\%$, compared
with the Asimov prediction of $3.62\%$.
\Cref{fig:asimov-prediction-vs-toys} shows the estimator and discovery
statistic distributions. The filled blue histograms, labelled ``Toys'',
show the hNDE pseudo-experiments. The black markers, labelled
``Simulator toys'', show a separate ensemble of $100{,}000$
pseudo-experiments generated with the original simulator at the same
parameter point. Both ensembles are analyzed with the same frozen hybrid
likelihood. The vertical error bars on the markers are the Poisson
statistical uncertainties of the simulator bin contents, divided by the
ensemble size to give probability per bin. The figure therefore tests both
the Asimov approximation and the agreement between the hNDE and
simulator ensembles. The implications for expected sensitivity are
discussed in \cref{sec:simulator-validation}.

\subsection{Integration precision with neural importance sampling}
\label{sec:demonstration-nis}

This test measures the integration gain from
Algorithm~\ref{alg:hNDE-nis-asimov}. We compare direct sampling from
$\qRef$ with NIS at equal sample sizes, using the same frozen hNDE
model and the same finite sample normalization procedure.

The proposal is trained on two million pilot points from $\qRef$.
The scan amplitude in \cref{eq:optimal-scan-proposal} is evaluated
at 13 equally spaced values over $0\leq\mu\leq3$, with equal
weights $a_k$. These weights are chosen before proposal training
and specify the relative priority of the design points. With this
choice, the formal proposal minimizes the leading contribution to
the mean variance of the test statistic across these points. Other
objectives can be encoded in the same weights: larger values can emphasize the null
hypothesis or anticipated confidence-interval crossings, while
weighting by the inverse square of a desired positive error tolerance
measures precision relative to that tolerance.

In the implementation, each $U_{\mu_k}$ is divided by the total
Asimov yield $\lambda(\boldsymbol{\theta}_A)$. This yield is fixed
at the generating point and is common to all scan points. The division
therefore multiplies $A$ by a single constant, which cancels in the
normalized proposal $g^\star$ and only rescales the weighted training
objective. It leaves the relative priorities unchanged and gives the
same formal proposal as using unscaled $U_{\mu_k}$ with $a_k=1$.
Weighted maximum likelihood is used to train the proposal flow.
The final defensive mixture uses $\epsilon=0.10$.
Architecture details and proposal-fit diagnostics are given in
App.~\ref{app:demonstration-training} and~\ref{app:demonstration-proposal}.

For each of the six sample sizes $M=2^9,\ldots,2^{14}$, we perform
128 independent repetitions and evaluate a 61-point scan over
$0\leq\mu\leq3$. Both component ratios are normalized on each
integration sample according to \cref{eq:nis-ratio-normalization}.
The maximum absolute score at $\mu_A$ is $1.9\times10^{-12}$,
including the samples with only $512$ events. Thus closure at the
generating point is retained as the integration sample is reduced.

We measure the spread of $q_{0,A}$ across repetitions by its
interquartile range (IQR) and define
\begin{equation}
  G=\left(\frac{\mathrm{IQR}_{\qRef}}
                 {\mathrm{IQR}_{\mathrm{NIS}}}\right)^2.
  \label{eq:nis-iqr-gain}
\end{equation}
The IQR is less sensitive to outliers than the sample variance.
For approximately Gaussian errors, $G$ estimates the variance ratio;
under the usual MC scaling, it also estimates the factor
by which direct sampling would need more events to achieve the same
precision. We separately assess the RMS error over the full scan
relative to the one million event benchmark.

\begin{figure}[htbp!]
  \centering
  \begin{subfigure}[t]{0.32\textwidth}
    \centering
    \includegraphics[width=\linewidth]
      {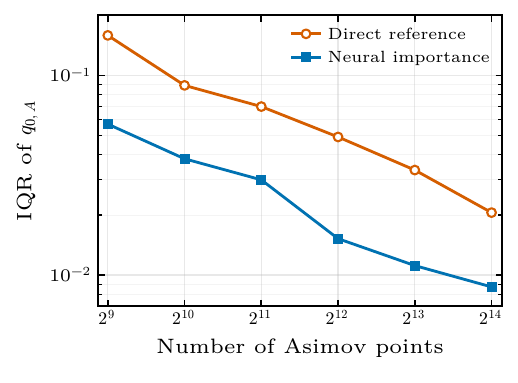}
    \caption{Discovery statistic.}
    \label{fig:nis-convergence-q0}
  \end{subfigure}
  \hfill
  \begin{subfigure}[t]{0.32\textwidth}
    \centering
    \includegraphics[width=\linewidth]
      {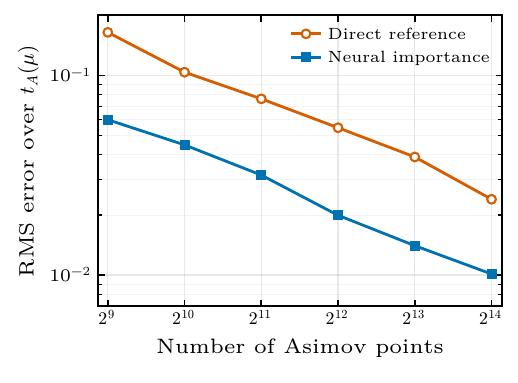}
    \caption{Complete likelihood scan.}
    \label{fig:nis-convergence-scan}
  \end{subfigure}
  \hfill
  \begin{subfigure}[t]{0.32\textwidth}
    \centering
    \includegraphics[width=\linewidth]
      {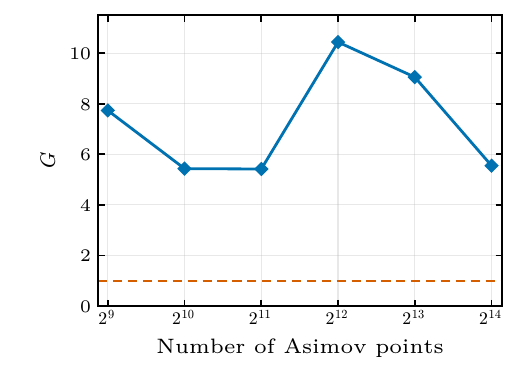}
    \caption{Squared IQR gain.}
    \label{fig:nis-event-saving}
  \end{subfigure}
  \caption{Comparison of sampling from the reference flow and NIS
at equal Asimov sample sizes, with 128 repetitions per size.
Panel~(\protect\subref{fig:nis-convergence-q0}) shows the IQR of
$q_{0,A}$; panel~(\protect\subref{fig:nis-convergence-scan}) shows
the RMS error over the full scan relative to the Asimov benchmark
constructed from one million events drawn from the reference flow; and
  panel~(\protect\subref{fig:nis-event-saving}) shows $G$.
  At $M=4{,}096$, the IQR decreases from $0.049$ to $0.015$ and the
  scan RMS error from $0.055$ to $0.020$.}
  \label{fig:nis-performance}
\end{figure}

At $M=4{,}096$, NIS reduces the IQR of $q_{0,A}$ from $0.049$ to
$0.015$, with $G=10.44$. The full-scan RMS error decreases from
$0.055$ to $0.020$. Across the tested sample sizes, $G$ ranges from
$5.4$ to $10.4$, as shown in \cref{fig:nis-performance}.
These comparisons quantify a reduction in the integration sample
needed for a given precision. Training and validation still require
large samples; the gain reduces the number of events evaluated in
subsequent likelihood calculations.

The practical size of the Asimov sample is particularly relevant for HEP
applications. With only $M=4{,}096$ weighted unbinned events, the full-scan
RMS error of $0.020$ is at the percent scale relative to order-one
test-statistic values. The RMS is expressed in test-statistic units;
pointwise relative precision depends on the scan position. The sample
contains more than three orders of magnitude fewer integration points
than the five-million-event reference used above.

For HEP signal regions containing a few thousand selected collision
events, an integration sample of this size would also be comparable in
event count to the data entering the observed likelihood. That comparison
depends on the analysis selection; the expected selected yield in this
toy model is approximately $153{,}434$. Reducing the integration sample
can lower the cost of every expected likelihood scan, NP
profile, and impact calculation. These repeated operations are a
substantial computational burden in the ATLAS NSBI
implementation~\cite{ATLAS:2024rpr}.

\subsection{Closure with a shape systematic uncertainty}
\label{sec:demonstration-systematics}

The final part of this demonstration checks the requirement to normalize the complete
parameter-dependent ratios when profiling a shape NP.
We introduce a detector-response parameter $\alpha_{\mathrm{scale}}$,
abbreviated as $\alpha$ below. The latent distributions are unchanged,
while the response means at the two variation points are shifted
coherently by $\pm10\%$:
\begin{equation}
  \mathbf{x}\mid\mathbf{z},\alpha=\pm1
  \sim\Normal\!\left(
    (1\pm0.1)\,\mathbf{z}\odot\mathbf{s},
    \operatorname{diag}(\boldsymbol{\sigma}^2)\right).
  \label{eq:scale-systematic-response}
\end{equation}
Here $\mathbf{s}$ contains the nominal response scales,
$\boldsymbol{\sigma}$ the smearing widths, and $\odot$ denotes
component-wise multiplication. A unit Gaussian constraint centered
at $\alpha=0$ is included in the likelihood. The varied samples are
normalized to the nominal selected yields, so this NP changes
only the shapes.

This differs from the signal strength $\mu$, which multiplies the complete
signal intensity and factorizes from the conditional process density.
When only normalization-varying parameters are present, the finite-sample normalization can be fixed as the parameters vary.
Shape factors, whether POIs or NPs, depend on both the event and $\alpha$, so the finite-sample normalization must be controlled throughout the NP scan. Shape factors are ubiquitous when describing experimental response and theoretical modeling uncertainties.

For each process, classifiers estimate the variation-to-nominal
density ratios
\begin{equation}
  v_s^\pm(\mathbf{x})
  \simeq\frac{p_s(\mathbf{x};\alpha=\pm1)}
               {p_s(\mathbf{x};\alpha=0)},
  \qquad s\in\{S,B\}.
  \label{eq:scale-systematic-anchors}
\end{equation}
We denote their smooth polynomial--exponential interpolation (as defined in HistFactory~\cite{Cranmer:1456844}) by $v_s(\mathbf{x};\alpha)$, with $v_s(\mathbf{x};0)=1$.
Normalizing the nominal and varied densities at the three input
points does not generally preserve normalization between them.
Let $r_{s,\boldsymbol{\psi}}(\mathbf{x};0)$ be the population-normalized
nominal sample-to-reference ratio of \cref{eq:hybrid-sample-density}. Preserving the
component normalization requires
\begin{equation}
  \E_{\qRef}\!\left[
    r_{s,\boldsymbol{\psi}}(\mathbf{x};0)v_s(\mathbf{x};\alpha)\right]=1.
  \label{eq:systematic-shape-normalization}
\end{equation}
The smooth polynomial--exponential interpolation reproduces the input anchors, but is nonlinear point by point and need not preserve the integral of a normalized shape. For comparison, piecewise-linear interpolation between normalized
density anchors preserves the integral within the anchor interval:
\begin{equation}
  v_s(\mathbf{x};\alpha)=
  \begin{cases}
    (1-\alpha)+\alpha v_s^+(\mathbf{x}), & 0\leq\alpha\leq1,\\
    (1+\alpha)-\alpha v_s^-(\mathbf{x}), & -1\leq\alpha<0.
  \end{cases}
  \label{eq:linear-normalized-systematic-interpolation}
\end{equation}
since each branch is a convex combination of normalized densities.

We therefore normalize the complete component ratio on the same
sample used to construct the Asimov weights:
\begin{align}
  Z_s(\alpha)
  &=\frac1M\sum_{m=1}^{M}
    r_{s,\boldsymbol{\psi}}(\mathbf{x}_m;\alpha),\notag\\
  \widetilde r_{s,\boldsymbol{\psi}}(\mathbf{x};\alpha)
  &=\frac{r_{s,\boldsymbol{\psi}}(\mathbf{x};\alpha)}{Z_s(\alpha)}
   =\frac{r_{s,\boldsymbol{\psi}}(\mathbf{x};0)v_s(\mathbf{x};\alpha)}
          {M^{-1}\sum_m r_{s,\boldsymbol{\psi}}(\mathbf{x}_m;0)
                            v_s(\mathbf{x}_m;\alpha)},
  \qquad \mathbf{x}_m\sim\qRef.
  \label{eq:alpha-dependent-shape-normalization}
\end{align}
This is the normalization of \cref{eq:sample-dependent-normalizer}
applied to the complete shape-dependent ratio. It gives
$M^{-1}\sum_m\widetilde r_{s,\boldsymbol{\psi}}
(\mathbf{x}_m;\alpha)=1$ at every evaluated NP value.
The process sum entering the finite Asimov likelihood is consequently
\begin{align}
  \widetilde h(\mathbf{x};\mu,\alpha)
  &=\sum_{s}\lambda_s(\mu)
    \widetilde r_{s,\boldsymbol{\psi}}(\mathbf{x};\alpha)=\mu\lambda_S
    \frac{r_{S,\boldsymbol{\psi}}(\mathbf{x};\alpha)}
         {Z_S(\alpha)}
    +\lambda_B
    \frac{r_{B,\boldsymbol{\psi}}(\mathbf{x};\alpha)}
         {Z_B(\alpha)}.
  \label{eq:demonstration-shape-model}
\end{align}

\begin{figure}[tbp!]
  \centering
  \begin{subfigure}[t]{0.49\textwidth}
    \centering
    \includegraphics[width=\textwidth]
      {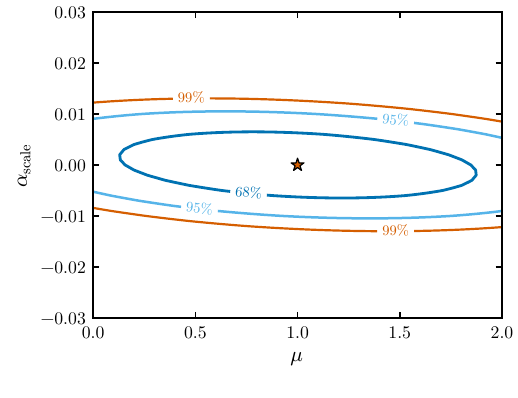}
    \caption{Two-dimensional likelihood.}
    \label{fig:shape-only-systematics-2d}
  \end{subfigure}
  \hfill
  \begin{subfigure}[t]{0.49\textwidth}
    \centering
    \includegraphics[width=\textwidth]
      {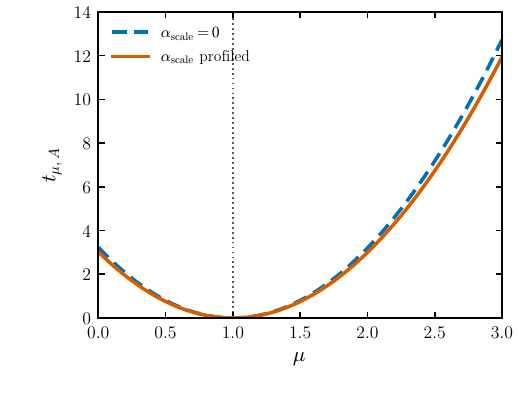}
    \caption{signal strength profile.}
    \label{fig:shape-only-systematics-profile}
  \end{subfigure}
  \caption{Asimov closure and NP profiling after complete
  shape-ratio normalization.
  Panel~(\protect\subref{fig:shape-only-systematics-2d}) shows the
  likelihood ratio statistic in the $(\mu,\alpha_{\mathrm{scale}})$
  plane. The contours use the $68\%$, $95\%$, and $99\%$
  two-parameter asymptotic thresholds; the star marks the coincident
  generating point and best fit.
  Panel~(\protect\subref{fig:shape-only-systematics-profile}) compares
  the signal strength scan with the NP fixed and profiled.}
  \label{fig:shape-only-systematics}
\end{figure}

Without this parameter-dependent normalization, the best fit is displaced from the
generating point $(\mu_A,\alpha_A)=(1,0)$ to approximately
$(0.635,0.0106)$. Complete normalization restores the best fit to
$(1,0)$. Profiling the NP then reduces the expected discovery
statistic from $3.230$ to $3.029$ while retaining the Asimov minimum
at the generating point, as shown in
\cref{fig:shape-only-systematics}.

This example demonstrates finite sample closure for a shape-dependent
likelihood as well as for a yield-only parameter. The result follows
from normalizing the complete ratios in every fitted parameter
direction on the same integration sample.

The need to recalculate the normalization increases the computational complexity of profiling the log likelihood. Using an interpolation that preserves the normalization is, therefore, advantageous in practical applications. An alternative approach, introduced in the ATLAS NSBI measurement~\cite{ATLAS:2024rpr}, interpolates the complete expression for the intensity. This option preserves the relation between the integrated intensity and the expected yield and, therefore, the finite-sample closure of the Asimov sample. More recently, flow-based models were proposed to model variations as parameter-dependent deformations of a nominal density~\cite{Valsecchi:2026flows}. The Jacobian in the flow transformation preserves continuous density normalization at each NP value.

The source code and notebooks for the demonstration are publicly
available~\cite{LopesDeSa:hnsbi-asimov}.

\section{Discussion and outlook}
\label{sec:discussion}
The contribution of hNDE is to connect classifier-based NSBI to an
evaluable density model and a reference from which additional events
can be generated on demand. The flow supplies this tractable reference,
while classifiers trained against its own generated samples learn
the target-to-reference ratios. With common support and exact
population ratios, their product with the reference recovers the
target densities without requiring the flow itself to model a
physical target. A reference that captures the dominant structures of
the targets can make the ratios easier to learn, although
classifiers still require validation. This extends
the ratio-based NSBI formalism implemented by
ATLAS~\cite{ATLAS:2024rpr} to a generative density model: the same
reference supports pseudo-experiment generation, weighted Asimov datasets whose generating point is an exact maximizer of the finite likelihood,
and methods, like NIS, designed to reduce the numerical cost of likelihood scans. The proposal optimization changes where
the model is evaluated while preserving the population likelihood
of the frozen hNDE model.

The practical objective is to combine accurate inference with
efficient density evaluation and sample generation. The relative
costs of these operations depend on the flow
architecture~\cite{Papamakarios:2019flows}. In hNDE, the reference
flow can be chosen for efficiency in both operations, provided it
adequately covers the targets, while the classifiers learn the
target-to-reference ratios. Absolute density evaluation
combines the flow density with the normalized ratios. For profile
likelihood ratios, however, the common factor $\qRef$ cancels,
leaving only the yields and ratios to be evaluated.

This construction addresses the operational limitations identified
in Ref.~\cite{ATLAS:2024tgo}. Independent reference events can be
generated on demand, so numerical integration for expected scans
and NP impacts is no longer limited to a fixed
simulation pool. Reweighting and resampling this reference supports
pseudo-experiment generation throughout the parameter region of
interest, enabling Neyman constructions for the learned model
without training a separate generator at every point. 

NIS further improves integration efficiency by concentrating events where their contributions to the numerical precision relevant to the measurement are greatest.  In the demonstration, the precision gains for $q_{0,A}$ correspond to estimated reductions by factors as large as 10 in the integration sample size needed for a given precision. This reduces the
number of events processed in subsequent likelihood calculations,
although training and validation still require large samples.
The total computing gain also depends on proposal training and
evaluation costs. The achievable reduction in more realistic
analyses remains to be established.

The following limitations govern the use and validation of hNDE.
\begin{itemize}
  \item \textit{Reference support and overlap.}
  The reference must cover every target throughout the parameter
  region used for inference. Where the reference density is much
  smaller than a target, large ratios can make classifier training
  difficult and allow a few events to dominate weighted samples.
  Training on a mixture of processes and relevant parameter
  variations, with suitable tail behavior, can improve overlap
  but does not guarantee it. Ratio tails and effective sample
  sizes should be monitored across the parameter region.

  \item \textit{Classifier accuracy and simulator sample validation.}
  Empirical normalization fixes the weighted average of each ratio
  on the reference sample, but it does not repair estimation errors.
  Calibration, reweighting checks, ensemble stability, and comparisons
  with independent simulator samples remain necessary. Exact Asimov
  closure and pseudo-experiments generated from the same learned
  model cannot by themselves establish accuracy or coverage under
  the simulator. Those claims require external validation and,
  where needed, calibration using simulator pseudo-experiments.

  \item \textit{Finite sample integration.}
  Under the positivity conditions of~\cref{sec:asimov}, normalizing the complete parameter dependent
  ratios on the same fixed weighted sample makes the generating
  point a global maximizer. This does not guarantee an accurate
  likelihood curvature or scan away from that point. Interpreting
  the Asimov result as median expected sensitivity also retains
  the usual asymptotic assumptions. Repeated integrations or block
  estimates should account for fluctuations of the normalizers.
  Comparisons between sampling from the reference and importance
  sampling, with the learned model fixed, help assess numerical
  uncertainty separately from model error.
\end{itemize}

Further studies should also assess whether a single NIS proposal provides
useful precision across several POI scans and NP
variations. Mixtures of flows or iterative proposal refinement could
be explored where a single flow provides insufficient overlap.

\subsection{Simulator validation of expected sensitivity}
\label{sec:simulator-validation}

Independent simulator validation provides the empirical basis for
using the hNDE Asimov dataset in an expected sensitivity study.
Exact Asimov closure establishes consistency with the learned
likelihood: under the conditions of \cref{sec:asimov}, the generating
point is a global maximizer even if the classifiers are
inaccurate~\cite{LopesDeSa:finite-asimov}. This checks the construction
and its normalization, but does not establish agreement with the
original simulator. Likewise, generating and fitting pseudo-experiments
with the same learned model cannot reveal errors shared by its
generation and inference procedures.

The external check is to generate separate ensembles from hNDE and
from the original simulator at the same parameter values, and analyze
both with the same frozen hNDE likelihood. Comparing the resulting distributions
of parameter estimators and test statistics then tests whether the
learned generator reproduces the behaviour of the analysis on
simulator data.

This comparison directly supports the use of the hNDE Asimov
prediction when the quantities relevant to the sensitivity study
agree to the required precision. The Asimov approximation must also
be checked against ensemble results: its interpretation as median
expected sensitivity relies on the usual asymptotic
approximations~\cite{Cowan:2010js}. Agreement between the Asimov
prediction and ensemble sensitivity, together with agreement between
the hNDE and simulator ensembles, therefore provides a practical
validation of the final sensitivity estimate.

For the demonstration in \cref{sec:demonstration}, the combined
comparison in \cref{fig:asimov-prediction-vs-toys} shows $100{,}000$
pseudo-experiments from each generator at $\mu_{\rm true}=1$, analyzed
with the same frozen hybrid likelihood. The distributions of
$\widehat\mu$ and $q_0$ agree within the statistical precision shown.
Their agreement with the Asimov predictions supports the sensitivity
estimate for the benchmark studied here.

A dense validation grid may be prohibitively expensive. Comparisons
at a limited set of representative POI and NP values,
including points where modelling is expected to be difficult, can
nevertheless provide meaningful confidence in using the hNDE Asimov
construction for the broader sensitivity study. These checks should
resolve discrepancies at the precision required by that study.
Their scope remains limited by the parameter points, distribution
regions, and ensemble sizes examined; agreement at selected points
does not establish accuracy everywhere.

Validation of expected sensitivity should also be distinguished from
calibration of confidence intervals. A Neyman construction calibrated
with independent simulator pseudo-experiments at the relevant
parameter points can provide valid coverage even when the learned
likelihood is imperfect~\cite{Brehmer:2018eca}, subject to the precision
of the calibration and appropriate treatment of NPs.
Calibration using only hNDE pseudo-experiments instead refers to the
learned model. Applying it to the simulator requires evidence that
the relevant test-statistic quantiles are adequately reproduced.
\subsection{Related work}
\label{sec:related-work}

The hNDE formalism extends the ratio-based NRE formalism to a generative NDE model. Following the ideas implemented by ATLAS~\cite{ATLAS:2024rpr}, process-to-reference density ratios share a common reference and are combined with event yields and NP dependence to construct the likelihood ratios used for inference. hNDE retains
this structure and supplies an explicit reference density that can
be evaluated and sampled. Multiplying the learned ratios by this
reference then gives absolute densities and supports event generation,
extending NRE to NDE.

The role of the flow $\qRef$ is to define the reference, rather than
to model a particular physical target. Ignoring this role,  the pure algebraic structure
\begin{equation}
\widehat{p}(\mathbf{x};\boldsymbol{\theta})
  =
  \qRef(\mathbf{x})\,
  r_{\boldsymbol{\psi}}(\mathbf{x};\boldsymbol{\theta}),
\end{equation}
has precedents in work on generative models. Grover et al.~\cite{Grover:2019lfiw}
use a learned generator as the reference and estimate the
target-to-reference density ratio with a classifier to reweight its samples
toward the target distribution. The same idea has been used many other contexts. In HEP, DCTRGAN~\cite{Diefenbacher:2020dctrgan}
uses related reweighting to improve generative calorimeter simulation.
These works aim to improve a generator trained to approximate the
target distribution. Although hNDE uses the same trivial algebraic factorization, the generative model and the density ratios have very different roles.

Related uses of flows and reweighting appear in Boltzmann
generators~\cite{noe2019boltzmanngeneratorssampling} and flow-based
sampling for fermionic lattice field theories~\cite{Albergo:2021bna}.
There, a specified physical distribution provides the weights or
Metropolis acceptance probabilities needed to obtain the target
distribution from flow proposals. For the simulator targets considered
here, the likelihood is intractable, and the target-to-reference ratios
are instead learned from simulation.

The NIS idea also has found many applications in the literature. In HEP, for instance, it has been used for efficient unfolding~\cite{Butter:2025spinup} and efficient phase-space sampling~\cite{Heimel:2022wy}. While the basic idea is the same, none of these previous work use it to improve the numerical integration with Asimov samples.%

\section{Conclusion}
\label{sec:conclusion}
We have presented hNDE, a framework for NSBI that combines expressive
NRE with a reference flow that can be evaluated
and sampled. The flow defines the reference hypothesis and is chosen
to cover the support of the target distributions of interest.
Classifiers trained against samples drawn from this flow learn
the target-to-reference density ratios. After
normalization, multiplying the reference density by a learned ratio
provides an evaluable target-density surrogate, with generation
through resampling of the weighted reference. The same learned model can therefore
support likelihood evaluation, pseudo-experiment generation, and
expected sensitivity calculations within an NSBI analysis.

For frequentist analyses of independent events, this representation
supports extended likelihoods, pseudo-experiments, and Neyman
constructions. We also construct weighted unbinned Asimov datasets
with exact finite sample closure. Normalizing the complete component
ratios at every parameter point on the same weighted sample makes
the generating point a global maximizer of the finite learned
likelihood, under the stated positivity and auxiliary conditions.
This result applies to both yield and shape parameters without
requiring differentiability.

A further contribution is a neural importance proposal designed
for the numerical integration of expected likelihood ratio scans.
Its objective accounts for both ordinary sampling fluctuations
and fluctuations of the component normalizations. Exact importance
weights preserve the integrals of the frozen hNDE model, while a
second flow concentrates evaluations where they can reduce numerical
error. In the demonstration, the variance reduction corresponds
to an estimated reduction by a factor of approximately five to ten
in the number of events required for a given test-statistic precision.

Independent simulator validation connects these constructions to
their intended use in sensitivity studies. The agreement of
estimator and test-statistic distributions in our benchmark provides
empirical support at the tested parameter point; comparisons at
representative POI and NP values provide a practical
basis for assessing wider applications. With this validation,
hNDE extends NRE into a reusable statistical
model that supports density evaluation, renewable event generation,
and efficient frequentist inference within one framework.

\paragraph*{Data availability statement}

The data that support the findings of this study are openly available at \url{https://github.com/rafaellopesdesa/hnsbi_asimov}. The repository contains the code to reproduce the event generation and analysis presented in the demonstration.

\paragraph*{Declaration of generative artificial intelligence (AI) use}

We acknowledge the use of ChatGPT and Claude throughout this project for assistance in scripting and for help editing the manuscript; the authors take full responsibility for the content of this manuscript.

\paragraph*{Disclosure statement} The authors report there are no competing interests to declare.

\section*{Acknowledgments}
The organizers of the 2026 ML4HEP School at the Tata Institute of Fundamental Research, where the initial idea for this project was developed, are gratefully acknowledged.  Support during the school was partially provided by the US National Science Foundation award HSF-India OISE-2201990. JS is also grateful for insightful discussions with Kyle Cranmer on the subject. The work of RCLSA is partially supported by the US Department of Energy award DE-SC0010004. The work of JS is partially supported by the US National Science Foundation IRIS-HEP Cooperative Agreement PHY-2323298.

\clearpage

\printbibliography

@misc{LopesDeSa:finite-asimov,
  author = {Lopes de S{\'a}, Rafael Coelho and Sandesara, Jay},
  title  = {Finite {Asimov} Sample Construction in Unbinned Neural Simulation-Based Inference},
  eprint = {2609.14136},
  archivePrefix = {arXiv},
  primaryClass = {physics.data-an},
  year   = {2026}
}

@article{Albergo:2021bna,
    author = "Albergo, Michael S. and Kanwar, Gurtej and Racani{\`e}re, S{\'e}bastien and Rezende, Danilo J. and Urban, Julian M. and Boyda, Denis and Cranmer, Kyle and Hackett, Daniel C. and Shanahan, Phiala E.",
    title = "{Flow-based sampling for fermionic lattice field theories}",
    eprint = "2106.05934",
    archivePrefix = "arXiv",
    primaryClass = "hep-lat",
    reportNumber = "MIT-CTP/5307",
    doi = "10.1103/PhysRevD.104.114507",
    journal = "Phys. Rev. D",
    volume = "104",
    number = "11",
    pages = "114507",
    year = "2021"
}

@misc{noe2019boltzmanngeneratorssampling,
      title={Boltzmann Generators -- Sampling Equilibrium States of Many-Body Systems with Deep Learning}, 
      author={Frank Noé and Simon Olsson and Jonas Köhler and Hao Wu},
      year={2019},
      eprint={1812.01729},
      archivePrefix={arXiv},
      primaryClass={stat.ML},
      url={https://arxiv.org/abs/1812.01729}, 
}

@article{Hesterberg01051995,
author = {Tim Hesterberg},
title = {Weighted Average Importance Sampling and Defensive Mixture Distributions},
journal = {Technometrics},
volume = {37},
number = {2},
pages = {185--194},
year = {1995},
publisher = {Taylor \& Francis},
doi = {10.1080/00401706.1995.10484303},
}

@misc{Cranmer:2015bka,
  author        = {Cranmer, Kyle and Pavez, Juan and Louppe, Gilles},
  title         = {Approximating Likelihood Ratios with Calibrated Discriminative Classifiers},
  eprint        = {1506.02169},
  archivePrefix = {arXiv},
  primaryClass  = {stat.AP},
  year          = {2015}
}

@ARTICLE{2020PNAS,
       author = {{Cranmer}, Kyle and {Brehmer}, Johann and {Louppe}, Gilles},
        title = "{The frontier of simulation-based inference}",
      journal = {Proceedings of the National Academy of Sciences},
         year = 2020,
        month = dec,
       volume = {117},
       number = {48},
        pages = {30055-30062},
          doi = {10.1073/pnas.1912789117},
archivePrefix = {arXiv},
       eprint = {1911.01429},
 primaryClass = {stat.ML},
       adsurl = {https://ui.adsabs.harvard.edu/abs/2020PNAS..11730055C}
}

@article{Brehmer:2018eca,
  author        = {Brehmer, Johann and Cranmer, Kyle and Louppe, Gilles and Pavez, Juan},
  title         = {Constraining Effective Field Theories with Machine Learning},
  journal       = {Phys. Rev. Lett.},
  volume        = {121},
  number        = {11},
  pages         = {111801},
  year          = {2018},
  doi           = {10.1103/PhysRevLett.121.111801},
  eprint        = {1805.00013},
  archivePrefix = {arXiv},
  primaryClass  = {hep-ph}
}

@article{Brehmer:2018hga,
  author        = {Brehmer, Johann and Cranmer, Kyle and Louppe, Gilles and Pavez, Juan},
  title         = {A Guide to Constraining Effective Field Theories with Machine Learning},
  journal       = {Phys. Rev. D},
  volume        = {98},
  number        = {5},
  pages         = {052004},
  year          = {2018},
  doi           = {10.1103/PhysRevD.98.052004},
  eprint        = {1805.00020},
  archivePrefix = {arXiv},
  primaryClass  = {hep-ph}
}

@article{ATLAS:2024rpr,
  author        = {{ATLAS Collaboration}},
  title         = {An implementation of neural simulation-based inference for parameter estimation in ATLAS},
  journal       = {Rep. Prog. Phys.},
  volume        = {88},
  number        = {6},
  pages         = {067801},
  year          = {2025},
  doi           = {10.1088/1361-6633/add370},
  eprint        = {2412.01600},
  archivePrefix = {arXiv},
  primaryClass  = {physics.data-an}
}

@article{ATLAS:2024tgo,
  author        = {{ATLAS Collaboration}},
  title         = "Measurement of off-shell Higgs boson production in the $H^{*}\to ZZ\to4\ell$ decay channel using a neural simulation-based inference technique in 13 TeV $pp$ collisions with the ATLAS detector",
  journal       = {Rep. Prog. Phys.},
  volume        = {88},
  number        = {5},
  pages         = {057803},
  year          = {2025},
  doi           = {10.1088/1361-6633/adcd9a},
  eprint        = {2412.01548},
  archivePrefix = {arXiv},
  primaryClass  = {hep-ex}
}

@inproceedings{Rezende:2015flows,
  author        = {Rezende, Danilo Jimenez and Mohamed, Shakir},
  title         = {Variational Inference with Normalizing Flows},
  booktitle     = {Proceedings of the 32nd International Conference on Machine Learning},
  series        = {Proceedings of Machine Learning Research},
  volume        = {37},
  pages         = {1530--1538},
  year          = {2015},
  eprint        = {1505.05770},
  archivePrefix = {arXiv},
  primaryClass  = {stat.ML}
}

@inproceedings{Durkan:2019splines,
  author        = {Durkan, Conor and Bekasov, Artur and Murray, Iain and Papamakarios, George},
  title         = {Neural Spline Flows},
  booktitle     = {Advances in Neural Information Processing Systems},
  volume        = {32},
  year          = {2019},
  eprint        = {1906.04032},
  archivePrefix = {arXiv},
  primaryClass  = {stat.ML}
}

@article{Papamakarios:2019flows,
  author        = {Papamakarios, George and Nalisnick, Eric and Rezende, Danilo Jimenez and Mohamed, Shakir and Lakshminarayanan, Balaji},
  title         = {Normalizing Flows for Probabilistic Modeling and Inference},
  journal       = {J. Mach. Learn. Res.},
  volume        = {22},
  number        = {57},
  pages         = {1--64},
  year          = {2021},
  eprint        = {1912.02762},
  archivePrefix = {arXiv},
  primaryClass  = {stat.ML}
}

@article{Cowan:2010js,
  author        = {Cowan, Glen and Cranmer, Kyle and Gross, Eilam and Vitells, Ofer},
  title         = {Asymptotic formulae for likelihood-based tests of new physics},
  journal       = {Eur. Phys. J. C},
  volume        = {71},
  pages         = {1554},
  year          = {2011},
  note          = {[Erratum: Eur. Phys. J. C 73, 2501 (2013)]},
  doi           = {10.1140/epjc/s10052-011-1554-0},
  eprint        = {1007.1727},
  archivePrefix = {arXiv},
  primaryClass  = {physics.data-an}
}

@article{Neyman:1937,
  author  = {Neyman, Jerzy},
  title   = {Outline of a Theory of Statistical Estimation Based on the Classical Theory of Probability},
  journal = {Philos. Trans. Roy. Soc. Lond. A},
  volume  = {236},
  pages   = {333--380},
  year    = {1937},
  doi     = {10.1098/rsta.1937.0005}
}

@misc{Valsecchi:2026flows,
  author        = {Valsecchi, Davide and Doneg\`a, Mauro and Wallny, Rainer},
  title         = {Factorizable Normalizing Flows for parameter-dependent density morphing},
  eprint        = {2606.30489},
  archivePrefix = {arXiv},
  primaryClass  = {hep-ex},
  year          = {2026}
}

@article{owen1997scrambled,
  author  = {Owen, Art B.},
  title   = {Scrambled Net Variance for Integrals of Smooth Functions},
  journal = {The Annals of Statistics},
  volume  = {25},
  number  = {4},
  pages   = {1541--1562},
  year    = {1997},
  doi     = {10.1214/aos/1031594731}
}

@misc{andral2024combining,
  author        = {Andral, Charly},
  title         = {Combining Normalizing Flows and Quasi-Monte Carlo},
  year          = {2024},
  eprint        = {2401.05934},
  archivePrefix = {arXiv},
  primaryClass  = {stat.CO},
  doi           = {10.48550/arXiv.2401.05934}
}

@misc{klebanov2023transporting,
  author        = {Klebanov, Ilja and Sullivan, T. J.},
  title         = {Transporting Higher-Order Quadrature Rules:
                   Quasi-Monte Carlo Points and Sparse Grids for
                   Mixture Distributions},
  year          = {2023},
  eprint        = {2308.10081},
  archivePrefix = {arXiv},
  primaryClass  = {math.NA},
  doi           = {10.48550/arXiv.2308.10081}
}

@article{muller2019neuralimportance,
  author        = {M{\"u}ller, Thomas and McWilliams, Brian and
                   Rousselle, Fabrice and Gross, Markus and
                   Nov{\'a}k, Jan},
  title         = {Neural Importance Sampling},
  journal       = {ACM Transactions on Graphics},
  volume        = {38},
  number        = {5},
  year          = {2019},
  eprint        = {1808.03856},
  archivePrefix = {arXiv},
  primaryClass  = {cs.LG},
  doi           = {10.1145/3341156}
}

@misc{Butter:2025spinup,
  author        = {Butter, Anja and Heimel, Theo and Huetsch, Nathan and Kagan, Michael and Plehn, Tilman},
  title         = {Simulation-Prior Independent Neural Unfolding Procedure},
  eprint        = {2507.15084},
  archivePrefix = {arXiv},
  primaryClass  = {hep-ph},
  year          = {2025}
}

@inproceedings{Hermans:2020,
  author    = {Hermans, Joeri and Begy, Volodimir and Louppe, Gilles},
  title     = "Likelihood-free MCMC with Amortized Approximate Ratio Estimators",
  booktitle = {Proceedings of the 37th International Conference on Machine Learning},
  series    = {Proceedings of Machine Learning Research},
  volume    = {119},
  pages     = {4239--4248},
  year      = {2020},
  eprint    = {1903.04057},
  archivePrefix = {arXiv},
  primaryClass  = {stat.ML},
  url       = {https://proceedings.mlr.press/v119/hermans20a.html}
}

@misc{LopesDeSa:hnsbi-asimov,
  author       = {Lopes de S{\'a}, Rafael Coelho and Sandesara, Jay},
  title        = {{hnsbi\_asimov}: Reproducible hNDE Asimov Likelihood Studies},
  year         = {2026},
  howpublished = {GitHub repository},
  url          = {https://github.com/rafaellopesdesa/hnsbi_asimov},
  urldate      = {2026-09-04}
}

@inproceedings{Grover:2019lfiw,
  author        = {Grover, Aditya and Song, Jiaming and Agarwal, Alekh and Tran, Kenneth and Kapoor, Ashish and Horvitz, Eric and Ermon, Stefano},
  title         = {Bias Correction of Learned Generative Models using Likelihood-Free Importance Weighting},
  booktitle     = {Advances in Neural Information Processing Systems},
  volume        = {32},
  year          = {2019},
  eprint        = {1906.09531},
  archivePrefix = {arXiv},
  primaryClass  = {stat.ML},
}

@article{Diefenbacher:2020dctrgan,
  author        = {Diefenbacher, Sascha and Eren, Engin and Kasieczka, Gregor and Korol, Anatolii and Nachman, Benjamin and Shih, David},
  title         = {{DCTRGAN}: Improving the Precision of Generative Models with Reweighting},
  journal       = {JINST},
  volume        = {15},
  number        = {11},
  pages         = {P11004},
  year          = {2020},
  doi           = {10.1088/1748-0221/15/11/P11004},
  eprint        = {2009.03796},
  archivePrefix = {arXiv},
  primaryClass  = {hep-ph}
}

@misc{Shapiro2000MonteCarlo,
  author       = {Shapiro, Alexander},
  title        = {Stochastic Programming by {Monte Carlo} Simulation Methods},
  year         = {2000},
  howpublished = {Stochastic Programming E-Print Series, No. 2000-3},
  doi          = {10.18452/8226},
  url          = {https://doi.org/10.18452/8226}
}

@techreport{Cranmer:1456844,
      author        = "Cranmer, Kyle and Lewis, George and Moneta, Lorenzo and
                       Shibata, Akira and Verkerke, Wouter",
      title         = "{HistFactory: A tool for creating statistical model for use with RooFit and RooStats}",
      institution   = "CERN",
      address       = "Geneva",
      number        = "CERN-OPEN-2012-016",
      month         = {1},
      year          = {2012},
      reportNumber  = "CERN-OPEN-2012-016",
      url           = "https://cds.cern.ch/record/1456844",
}

@article{Heimel:2022wy,
    author = "Heimel, Theo and Winterhalder, Ramon and Butter, Anja and Isaacson, Joshua and Krause, Claudius and Maltoni, Fabio and Mattelaer, Olivier and Plehn, Tilman",
    title = "{MadNIS - Neural multi-channel importance sampling}",
    eprint = "2212.06172",
    archivePrefix = "arXiv",
    primaryClass = "hep-ph",
    reportNumber = "IRMP-CP3-22-56, MCNET-22-22, FERMILAB-PUB-22-915-T",
    doi = "10.21468/SciPostPhys.15.4.141",
    journal = "SciPost Phys.",
    volume = "15",
    number = "4",
    pages = "141",
    year = "2023"
}
\clearpage
\appendix

\section{Mathematical details of importance sampling}
\label{app:asimov}
This appendix gives the mathematical details supporting
\cref{sec:efficient-asimov}: the statistical properties under importance
sampling with common normalization, the optimal-proposal argument and
weighted training objective, and the extension to a normalized NDE model.
We use the population quantities and statistical results defined in
\cref{sec:asimov} and keep the trained model and generating point fixed
throughout.

\subsection{Statistical properties under importance sampling}
\label{app:importance-normalization}

For a fixed normalized proposal $g$ whose support covers $\qRef$, the same
population integral is
\begin{equation}
  I_{\boldsymbol{\theta}}
  =\E_g\!\left[
      \frac{\qRef(\mathbf{x})}{g(\mathbf{x})}
      Y_{\boldsymbol{\theta}}(\mathbf{x})\right].
  \label{eq:app-asimov-importance}
\end{equation}
For learned proposals, the statements below are conditional on their
training sample; the final integration sample is independent of that sample.
With fixed component normalizers and raw importance weights
$\rho=\qRef/g$, the pointwise estimators remain unbiased and
\begin{equation}
  \Var[-2\widehat{\Delta\ell}^{\mathrm{fixed},g}]
     =\frac4M\Var_g[\rho Y_{\boldsymbol{\theta}}].
  \label{eq:importance-fixed-moments}
\end{equation}
Self-normalizing these weights alone would define different ratio
estimators. We now propagate self-normalized reference weights together
with the component normalizers, as required for finite Asimov closure.

Draw $\mathbf{x}_m\sim g$ independently. With
$\rho_m=\qRef(\mathbf{x}_m)/g(\mathbf{x}_m)$ and
$\omega_m=\rho_m/\sum_j\rho_j$, define
\begin{equation}
  S_0=\frac1M\sum_m\rho_m,
  \qquad
  B_{s,v}=\frac1M\sum_m\rho_m r_{s,v}(\mathbf{x}_m),
  \qquad v=A,\boldsymbol{\theta}.
  \label{eq:importance-normalizer-averages}
\end{equation}
Then $Z_{s,v}=B_{s,v}/S_0$.
Writing $H_v=\sum_s\lambda_{s,v}r_{s,v}/B_{s,v}$ gives
\begin{equation}
  \begin{aligned}
    \widetilde h_v&=S_0H_v,
    \qquad w_m^A=\frac{\rho_m}{M}H_A(\mathbf{x}_m),\\
    \widehat I_{\boldsymbol{\theta}}
    &=\frac1M\sum_m\rho_m H_A(\mathbf{x}_m)
      \log\frac{H_A(\mathbf{x}_m)}
                   {H_{\boldsymbol{\theta}}(\mathbf{x}_m)}.
  \end{aligned}
  \label{eq:importance-common-normalization}
\end{equation}
The common factor $S_0$ cancels from both the Asimov weights and the
ratio inside the logarithm.

The coefficients in \cref{eq:normalization-coefficients} satisfy
\begin{equation}
  \sum_s b^A_{s,\boldsymbol{\theta}}
  =I_{\boldsymbol{\theta}}+\lambda_A,
  \qquad
  \sum_s b^T_{s,\boldsymbol{\theta}}=\lambda_A.
  \label{eq:normalization-coefficient-sums}
\end{equation}
Expanding \cref{eq:importance-common-normalization} around $B_{s,v}=1$
and using these identities gives
\begin{equation}
  -2\widehat{\Delta\ell}(\boldsymbol{\theta})
  +2\Delta\ell(\boldsymbol{\theta})
  =\frac2M\sum_m\rho_m U_{\boldsymbol{\theta}}(\mathbf{x}_m)
    +o_p(M^{-1/2}),
  \label{eq:importance-asimov-expansion}
\end{equation}
under the corresponding moment and differentiability conditions for
importance sampling. Since $\E_g[\rho U_{\boldsymbol{\theta}}]=0$, the leading variance is
\begin{align}
  \Var_{\mathrm{MC}}[-2\widehat{\Delta\ell}(\boldsymbol{\theta})]
    &=\frac4M\mathcal \E_g[(\rho U_{\boldsymbol{\theta}})^2]+o(M^{-1}),\notag\\
  \mathcal 
    \E_g[(\rho U_{\boldsymbol{\theta}})^2]
     &=\int\frac{\qRef^2(\mathbf{x})U_{\boldsymbol{\theta}}^2(\mathbf{x})}
                  {g(\mathbf{x})}\dd\mathbf{x}.
  \label{eq:app-importance-second-moment}
\end{align}

These statements require the corresponding weighted moment, derivative,
and remainder controls. Under second-order expectation control, the
same-sample log likelihood and statistic have biases of order $M^{-1}$.
For the profiled statistic, evaluate $U$ at
$\boldsymbol{\theta}^{*}_{\boldsymbol{\mu}}$ and apply
\cref{sec:asimov-profiling}. Setting $g=\qRef$ recovers the direct-sampling
results. This derivation accounts for both the normalized reference weights
and the component normalizers. Section~\ref{sec:efficient-asimov} uses the
resulting variance coefficient to design an efficient sampling proposal
for discovery, exclusion, and confidence-interval scans.
\subsection{Optimal proposals and weighted likelihood training}
\label{app:nis-proposals}

For a single tested point, Cauchy--Schwarz gives
\begin{equation}
  \left(\int\qRef|U_{\boldsymbol{\theta}}|\dd\mathbf{x}\right)^2
  \leq
  \left(\int\frac{\qRef^2 U_{\boldsymbol{\theta}}^2}{g}
     \dd\mathbf{x}\right)
  \left(\int g\dd\mathbf{x}\right).
  \label{eq:importance-cauchy-schwarz}
\end{equation}
For normalized $g$, equality gives
\cref{eq:optimal-point-proposal} when its normalizing constant is
finite and positive. Replacing $|U_{\boldsymbol{\theta}}|$ by the
scan amplitude $A$ gives
\cref{eq:optimal-scan-proposal}. The objective is a weighted sum of
variances at the design points; it does not directly minimize the
maximum error across a scan or the variance of an interval crossing.
The defensive mixture in \cref{eq:defensive-mixture} restores support
where the formal proposal vanishes.

The weighted training objective satisfies
\begin{equation}
  \E_{\qRef}[A\log g_{\boldsymbol{\eta}}]
  =\E_{\qRef}[A]\,
     \E_{g^\star}[\log g_{\boldsymbol{\eta}}],
  \label{eq:nis-forward-kl}
\end{equation}
so maximizing it minimizes the KL divergence
$D_{\mathrm{KL}}(g^\star\Vert g_{\boldsymbol{\eta}})$.
Within a restricted flow family, this need not minimize the integration
variance~\cite{muller2019neuralimportance}. The fitted flow and its
defensive mixture therefore approximate the formal optimum; the
achieved variance reduction must be assessed on fresh integration
samples.

\subsection{Application to a normalized NDE model}
\label{app:nde-importance}

The construction also applies to a frozen, normalized and evaluable
NDE model
$\nu(\mathbf{x};\boldsymbol{\theta})=
\lambda(\boldsymbol{\theta})p(\mathbf{x};\boldsymbol{\theta})$.
For $\lambda_A>0$, choose $p_A=p(\cdot;\boldsymbol{\theta}_A)$ as
the reference and make the substitutions
\begin{equation}
  \qRef\longrightarrow p_A,
  \qquad
  Y_{\boldsymbol{\theta}}(\mathbf{x})
  \longrightarrow\lambda_A
    \log\frac{\nu(\mathbf{x};\boldsymbol{\theta}_A)}
                 {\nu(\mathbf{x};\boldsymbol{\theta})}.
  \label{eq:nde-importance-identification}
\end{equation}
Their product leaves the expected integral unchanged. For normalization
on the common sample, $p_A$ must cover every tested density; otherwise
a broader reference is required. Treating the total density as one
component, \cref{eq:normalization-influence} applies with
$r_{\boldsymbol{\theta}}=p(\cdot;\boldsymbol{\theta})/p_A$ and $r_A=1$.

Pilot events can be drawn from $p_A$ when a sampler is available, or
from another proposal with the corresponding importance weights.
Evaluable densities are required to compute those weights. This
extension improves integration of the NDE model; it does not correct
errors in the learned density.

\FloatBarrier

\section{Additional details of the demonstration}
\label{app:demonstration}
\subsection{Network architectures and training}
\label{app:demonstration-training}

\begin{figure}[htbp!]
  \centering
  \includegraphics[width=0.98\textwidth]
    {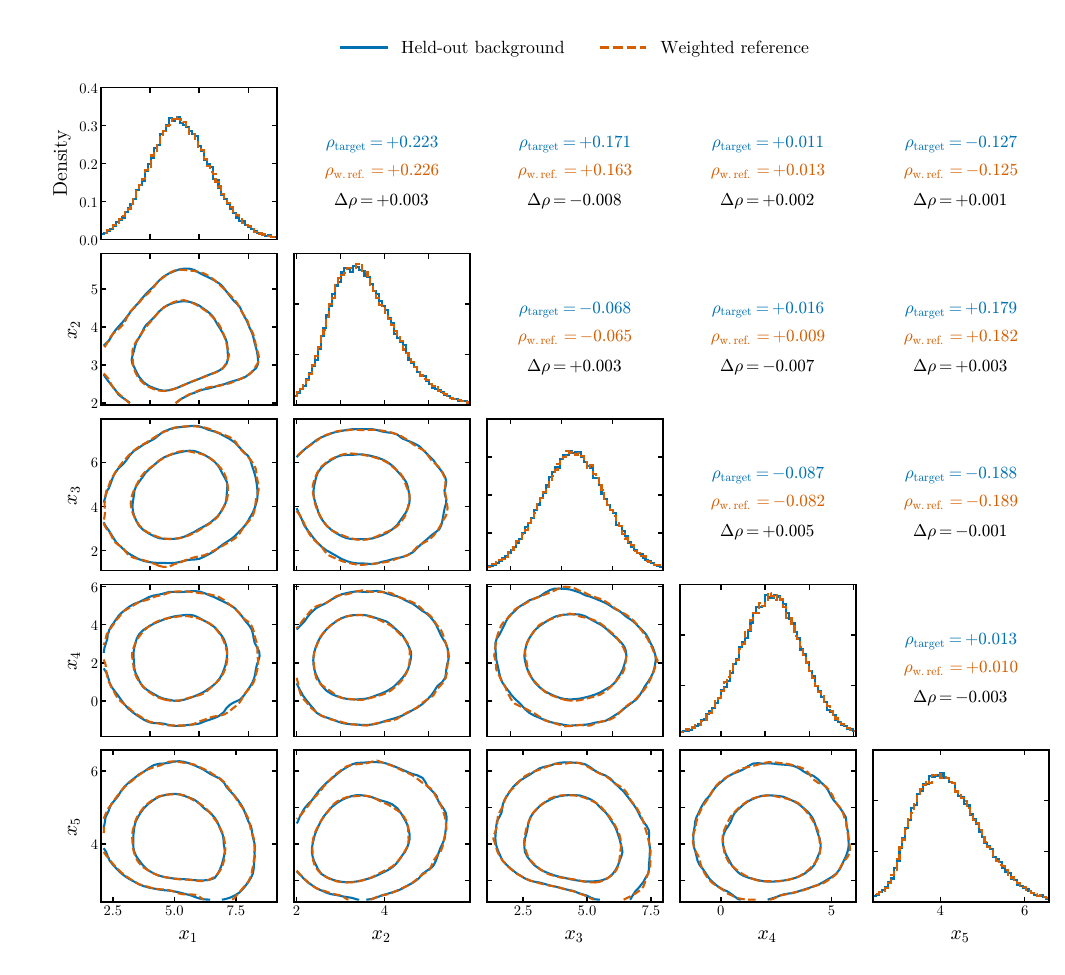}
  \caption{Background reweighting validation after preselection.
  The diagonal panels compare normalized one-dimensional densities
  from $100{,}000$ independent selected background events with an
  equally sized subsample of the five-million-event flow reference,
  reweighted by $r_{B,\boldsymbol{\psi}}$. The lower-triangle panels
  compare pairwise density contours, and the upper-triangle entries
  compare correlation coefficients. The distributions are normalized
  separately to test shape agreement.}
  \label{fig:hybrid-closure}
\end{figure}

The preselection classifier has three hidden layers of 256 nodes.
Its threshold on the estimated signal-to-background density ratio
is chosen to give an expected selected background-to-signal ratio
of approximately $250$.

The reference density uses a quadratic-spline normalizing
flow~\cite{Durkan:2019splines} with ten coupling transforms.
Each conditioner has four hidden layers of $1{,}024$ features.
The splines have 16 bins over the standardized interval $[-5,5]$
and linear tails outside it. The flow is trained by unweighted
maximum likelihood on $50{,}000$ selected signal and $50{,}000$
selected background events, with batches of $2{,}048$, an initial
learning rate of $10^{-4}$, and at most 70 epochs with early stopping.

Separate ensembles estimate the signal-to-reference and
background-to-reference ratios. Each ensemble contains four
independently initialized dense networks with four hidden layers of
$1{,}024$ nodes. Training uses five million events in each class, at most
50 epochs, batches of $1{,}024$, and an initial learning rate of
$10^{-3}$. Class weights are normalized separately and no
post-training calibration is applied. The arithmetic mean of the
four ratio estimates is used in the likelihood. Empirical
normalization is then performed on the integration sample used
for the corresponding calculation.

The NIS proposal uses a quadratic-spline flow with 12 coupling
transforms, four hidden layers of $1{,}024$ nodes, 24 spline bins, and
a tail bound of six. It is trained by weighted maximum likelihood
on the two-million-event pilot sample described in
\cref{sec:demonstration-nis}.
\subsection{Background reweighting checks}
\label{app:demonstration-reweighting}

\Cref{fig:hybrid-closure} tests whether the background ratio
reweights the reference to the selected background distribution.
The one-dimensional projections and pairwise comparisons assess
shape agreement independently of the physical signal and background
yields. The likelihood comparison in
\cref{sec:demonstration-accuracy} tests the effect of the learned
model on inference.

\subsection{Proposal-fit diagnostics}
\label{app:demonstration-proposal}

\begin{figure}[htbp!]
  \centering
  \includegraphics[width=0.5\textwidth]
    {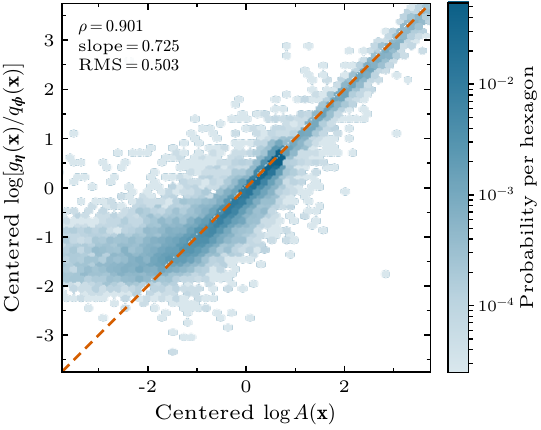}
  \caption{Proposal training diagnostic. The centered learned log
  density ratio $\log(g_{\boldsymbol{\eta}}/\qRef)$ is compared
  with the centered target $\log A$. The color scale gives the
  fraction of reference events in each hexagon on a logarithmic
  scale. The dashed diagonal marks agreement. The correlation,
  slope, and RMS characterize the proposal fit; the integration
  gain is measured separately in \cref{fig:nis-performance}.}
  \label{fig:nis-proposal}
\end{figure}

\Cref{fig:nis-proposal} compares the learned log density ratio
$\log(g_{\boldsymbol{\eta}}/\qRef)$ with the target $\log A$
after subtracting their respective means. The correlation is
$0.901$, with slope $0.725$ and RMS $0.503$. The learned proposal
therefore approximates the target without reproducing it exactly.
After importance reweighting, closure is obtained in all five
reference projections. The exact importance weights account for
the proposal density actually used, so proposal-fit errors affect
integration efficiency without changing the target integral.

\subsection{Performance comparison with flow-only estimation}
\label{app:flow-only-likelihood}

\begin{figure}[htbp!]
    \centering
    \includegraphics[width=0.5\linewidth]
        {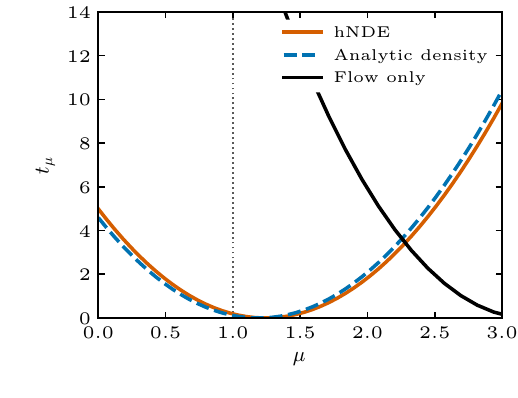}
    \caption{Likelihood ratio scans using hNDE, analytic densities,
    and flow-only density estimation (black). The vertical line marks the generating
    value $\mu=1$.}
    \label{fig:flow-only-profile-comparison}
\end{figure}

\Cref{fig:flow-only-profile-comparison} compares hNDE, analytic
truth, and a likelihood built from separately trained signal and
background flows. All three are evaluated on the dataset used
in \cref{fig:profile-hybrid-truth}. The scan using flows alone
continues to a minimum near $\mu=3$, substantially displaced from
the analytic fit. This illustrates how errors in the learned
densities can affect inference in this example.

The signal and background densities each use an ensemble of four
independently trained rational-quadratic-spline flows, trained on
the same data as the hNDE classifiers. Each flow has ten coupling
transforms with eight spline bins and linear tails outside
$[-5,5]$ in standardized coordinates. Each transform has a
conditioner containing four residual blocks with 1024 hidden
features. An individual flow is therefore substantially larger
than an individual ratio classifier, which has four hidden layers
of 1024 nodes.

\subsection{Effect of model misspecification}
\label{app:controlled-misspecification}

We introduce a controlled model error by replacing each normalized
process-to-reference ratio with a mixture of $95\%$ of itself and $5\%$ of the
other process-to-reference ratio. At $\mu_{\rm true}=1$, pseudo-experiments from
this deformed hNDE model and the original simulator are both
analyzed with the deformed likelihood.

\begin{figure}[htbp!]
    \centering

    \begin{subfigure}[t]{0.48\textwidth}
        \centering
        \includegraphics[width=\linewidth]
            {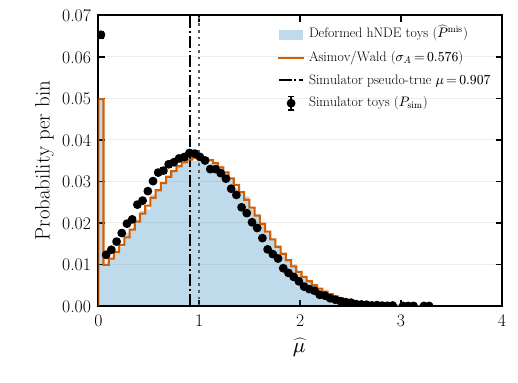}
        \caption{Signal-strength estimator.}
        \label{fig:misspecification-estimator}
    \end{subfigure}
    \hfill
    \begin{subfigure}[t]{0.48\textwidth}
        \centering
        \includegraphics[width=\linewidth]
            {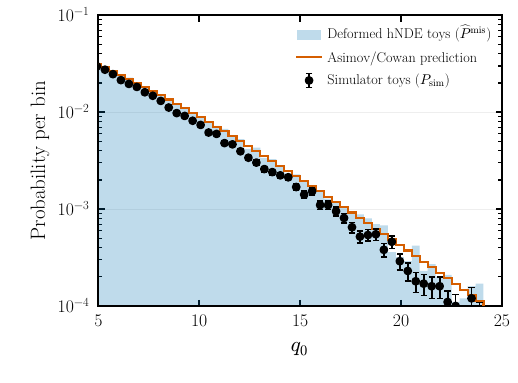}
        \caption{Discovery-statistic tail.}
        \label{fig:misspecification-discovery}
    \end{subfigure}

    \caption{Controlled model misspecification with a $5\%$ ratio
    deformation. Panels (\protect\subref{fig:misspecification-estimator})
    and (\protect\subref{fig:misspecification-discovery}) compare the
    estimator and discovery-statistic distributions with the Asimov
    predictions. Both toy ensembles are analyzed with the same deformed
    likelihood.}
    \label{fig:controlled-misspecification}
\end{figure}

In \cref{fig:misspecification-estimator,fig:misspecification-discovery},
the deformed hNDE ensemble (blue) follows its Asimov prediction (orange),
while the simulator ensemble (black, with Poisson statistical uncertainties)
reveals differences in the estimator distribution and the $q_0$ tail.
The latter changes the $p$-values inferred from the learned Asimov
prediction.
This illustrates why agreement within the learned model must be
complemented by independent simulator validation before using its
Asimov predictions for sensitivity studies. In the estimator panel,
the dotted line marks
$\mu_{\rm true}=1$; the dash-dotted line marks the signal strength
that maximizes the expected deformed log likelihood for simulator
data.

\FloatBarrier

\section{Supplemental material}
\begin{figure}[htbp!]
    \centering
    \includegraphics[width=\linewidth]{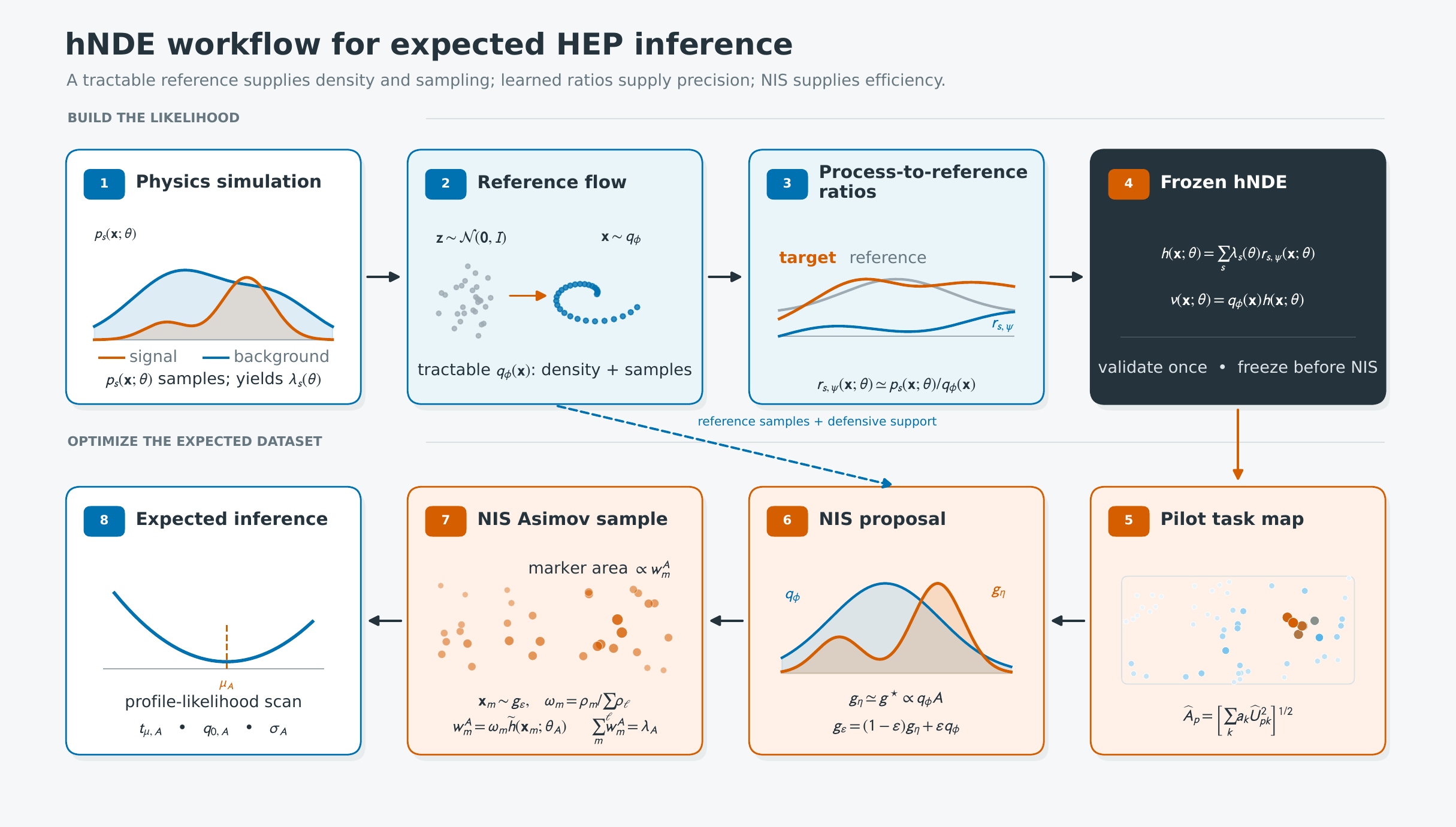}
    \caption{Overview of the hNDE workflow for expected HEP inference. Top: samples from the physics simulator are used to train a tractable reference flow $\qRef(\mathbf{x})$ and process-to-reference density ratios $r_{s,\boldsymbol{\psi}}(\mathbf{x};\boldsymbol{\theta})\simeq p_s(\mathbf{x};\boldsymbol{\theta})/\qRef(\mathbf{x})$. Bottom: pilot estimates $\widehat U_{pk}$ of the event fluctuations, including same-sample normalization, define the proposal-training weights $\widehat A_p$. The NIS flow $g_{\boldsymbol{\eta}}$ approximates the optimal proposal $g^\star\propto\qRef A$ and enters the defensive mixture $g_\epsilon=(1-\epsilon)g_{\boldsymbol{\eta}}+\epsilon\qRef$. Fresh samples from $g_\epsilon$ use self-normalized reference weights $\omega_m=\rho_m/\sum_\ell\rho_\ell$, with $\rho_m=\qRef(\mathbf{x}_m)/g_\epsilon(\mathbf{x}_m)$. Normalizing the component ratios on these same events gives $\widetilde h$ and the Asimov weights $w_m^A=\omega_m\widetilde h(\mathbf{x}_m;\boldsymbol{\theta}_A)$, which satisfy $\sum_m w_m^A=\lambda_A$ exactly. The resulting weighted Asimov dataset is used for the expected profile likelihood-ratio scan.}
    \label{fig:hNDE_diagram}
\end{figure}

\end{document}